\documentclass[aps,prd,superscriptaddress,preprintnumbers,reprint,nofootinbib,showpacs]{revtex4-1}
\pdfoutput=1

\usepackage{graphicx}
\usepackage{epsfig}
\usepackage{amssymb}

\def\als{\alpha_s} 
\def\MS{\overline{\rm MS}}
\newcommand{\be}{\begin{equation}}
\newcommand{\ee}{\end{equation}}
\newcommand{\bea}{\begin{eqnarray}}
\newcommand{\eea}{\end{eqnarray}}

\begin{document}

\title{Determination of $\alpha_s$ from the QCD static energy: an update}
\author{Alexei Bazavov}
\altaffiliation[Current address: ]{Department of Physics and Astronomy, University of Iowa, 203 Van Allen Hall, Iowa City, Iowa 52242-1479}
\affiliation{Physics Department, Brookhaven National Laboratory,
  Upton, NY 11973, USA}
\author{Nora Brambilla}
\affiliation{Physik Department, Technische Universit\"at M\"unchen, D-85748 Garching, Germany}
\affiliation{Institute for Advanced Study, Technische Universit\"at M\"unchen, Lichtenbergstrasse 2 a, D-85748 Garching, Germany}
\author{Xavier \surname{Garcia i Tormo}}
\affiliation{Albert Einstein Center for Fundamental Physics. Institut f\"ur Theoretische Physik, Universit\"at Bern,
  Sidlerstrasse 5, CH-3012 Bern, Switzerland}
\author{P\'eter Petreczky}
\affiliation{Physics Department, Brookhaven National Laboratory,
  Upton, NY 11973, USA}
\author{Joan Soto}
\affiliation{Departament d'Estructura i Constituents de la Mat\`eria and Institut de Ci\`encies del Cosmos, 
Universitat de Barcelona, Diagonal 647, E-08028 Barcelona, Catalonia, Spain}
\author{Antonio Vairo}
\affiliation{Physik Department, Technische Universit\"at M\"unchen, D-85748 Garching, Germany}

\date{\today}

\preprint{TUM-EFT 47/14, ~UB-ECM-PF 14/81, ~ICCUB-14-055}

\begin{abstract}
We present an update of our determination of the strong coupling $\alpha_s$ from the quantum chromodynamics static energy. This updated analysis includes new lattice data, at smaller lattice spacings and reaching shorter distances, the use of better suited perturbative expressions to compare with data in a wider distance range, and a comprehensive and detailed estimate of the error sources that contribute to the uncertainty of the final result. Our updated value for $\als$ at the $Z$-mass scale, $M_Z$, is $\alpha_s(M_Z)=0.1166^{+0.0012}_{-0.0008}$, which supersedes our previous result.
\end{abstract}


\maketitle

\section{Introduction}\label{sec:intr}
The strong coupling, $\alpha_s$, is the only free parameter of quantum chromodynamics (QCD) in the massless quark limit. The knowledge of its value with good precision and accuracy is of vital importance for the study of any process that involves the strong interactions~\cite{Brambilla:2014aaa}. It has, for instance, a big impact on cross-section calculations for the Large Hadron Collider, where its uncertainty is often one of the limiting factors on the precision reached by theoretical predictions. The present world average by the Particle Data Group (PDG)~\cite{Beringer:1900zz} quotes an uncertainty of about $\pm0.5\%$ for $\alpha_s$ at the scale of the $Z$-boson mass, $M_Z$. But, when one looks at the individual values that enter in the average, one realizes that the uncertainties of many of the individual measurements are dominated by errors of theoretical origin. These errors are often difficult to precisely assess, and in fact some of the individual measurements are not in good agreement with each other. A flagrant example of that is given, for instance, by comparing the lattice result obtained in Ref.~\cite{McNeile:2010ji} with the value obtained from analyses of the event-shape variable thrust in $e^+e^-$ collisions of Ref.~\cite{Abbate:2010xh}. Disagreements like that question, to some extent, if the uncertainty quoted in the world average really reflects our current understanding of the value of $\alpha_s$. In that sense it is crucial to have an increasing corroboration of the value of $\alpha_s$, by extracting it from different independent quantities, and at different energy ranges, and to critically and exhaustively analyze all the theoretical errors that enter in each of the determinations. In Ref.~\cite{Bazavov:2012ka} we presented a novel determination of $\alpha_s$, based on the comparison of lattice data for the energy between two static sources in the fundamental representation of QCD, i.e. the QCD static energy, at short distances with perturbative expressions. This extraction was possible due to recent progress both in the lattice evaluation and the perturbative computation of the static energy. It provided a competitive determination of $\alpha_s$, that stemmed from a perturbative calculation at three-loop order, including resummations of logarithmically enhanced terms. In this paper we present an update of the result of Ref.~\cite{Bazavov:2012ka}. The main new ingredients are: (i) we include new lattice data, with smaller lattice spacings, and reaching shorter distances, (ii) we use perturbative expressions that are better suited to compare with lattice data in this wider distance range, and (iii) we include a more detailed assessment of the different errors sources that can contribute to the uncertainty of the final result; the discussion of the different errors sources comprises, in fact, the bulk of the present paper.

The rest of the paper is organized as follows. In Sec.~\ref{sec:lattice} we describe the lattice data we use, in particular explaining in detail the lattice systematic errors at short distances. Section~\ref{sec:pert} displays the perturbative expressions we employ. The bulk of the paper is contained in Sec.~\ref{sec:asextr}, where we explain in detail the analysis to extract $\alpha_s$, and discuss all the error sources that can affect our result. Section~\ref{sec:disc} contains some discussion of our results, as well as comparison with previous related works. Finally, in Sec.~\ref{sec:concl} we present a short summary of the main results and conclude.

\section{Lattice data}\label{sec:lattice}
\begin{figure}
\includegraphics[width=8.6cm]{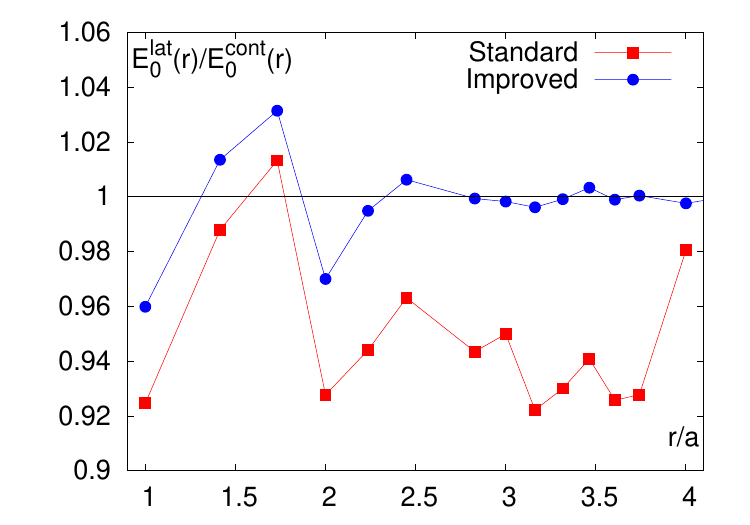}
\caption{The ratio of the lattice and the continuum static static energy at tree level.} 
\label{fig:free}
\end{figure}

\begin{figure}
\includegraphics[width=8.6cm]{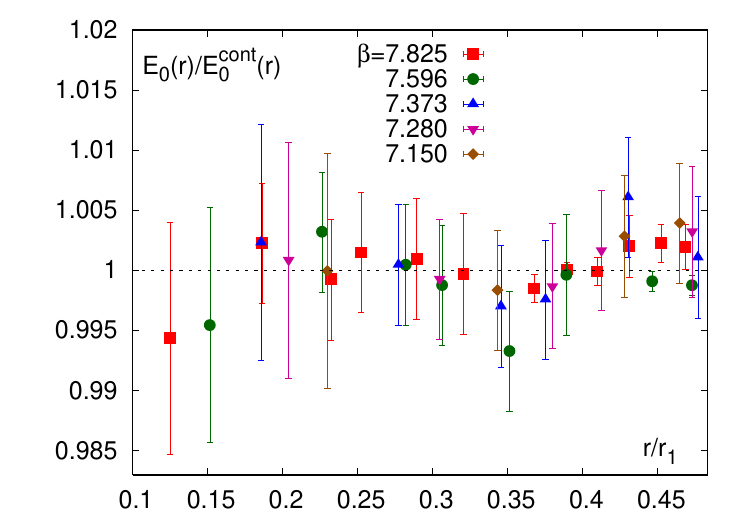}
\caption{The ratio of the corrected static energy at $\beta=7.150,~7.280,~7.373,~7.596$ and $7.825$ to the continuum estimate.}
\label{fig:ratall}
\end{figure}

In this paper we use 2+1-flavor lattice QCD data on the static quark-antiquark energy, $E_0(r)$, obtained using tree-level improved gauge action and 
Highly-Improved Staggered Quark (HISQ) action, procured by the HotQCD collaboration~\cite{Bazavov:2014noa}. The strange-quark mass $m_s$ 
was fixed to its physical value, while the light-quark masses were
chosen to be $m_l=m_s/20$. These correspond to a pion mass of about 
$160$ MeV in the continuum limit, which is very close to the physical value. More precisely, we use lattice QCD data corresponding
to the lattice gauge couplings $\beta=10/g^2=7.150,~ 7.280, 7.373, 7.596$ and $7.825$. The largest gauge coupling, $\beta=7.825$, corresponds to lattice spacings
of $a=0.041$~fm. One may worry about the evolution of the topological charge on such fine lattices, but, as it was demonstrated in Ref.~\cite{Bazavov:2014noa}, the Monte-Carlo evolution of the topological charge is acceptable even for $\beta=7.825$. The lattice spacing was fixed using the $r_1$ scale defined as
\begin{equation}
\left.r^2 \frac{d E_0(r)}{ d r}\right|_{r=r_1}=1.0\,.
\end{equation}
The values of $r_1/a$ for each of the above gauge couplings have been
determined in Ref.~\cite{Bazavov:2014noa} and are summarized in Tab.~\ref{tab:r1}. The errors on the values of $r_1/a$ in the table are the combined statistical and systematic errors. The details of the lattice spacing determination, including the error estimates, are presented in Ref.~\cite{Bazavov:2014noa}, where
also additional checks on the lattice spacing determination using different physical observables are performed. Since we are interested in the behavior of the static energy at short distances the effect of finite volume in lattice calculations is negligible.
\begin{table*}
\begin{tabular}{|c|ccccc|}
\hline
$\beta$ & 7.150     & 7.280     & 7.373     & 7.596     & 7.825     \\
\hline
$r_1/a$ & 4.212(42) & 4.720(33) & 5.172(34) & 6.336(56) & 7.690(58) \\
\hline
Volume & $48^3\times64$ & $48^3\times64$ & $48^3\times64$ & $64^4$ & $64^4$\\
\hline
\end{tabular}
\caption{The values of $r_1/a$ and the lattice volumes for the different gauge couplings considered in this paper.
The errors are the combined statistical and systematic errors.}
\label{tab:r1}
\end{table*}
However lattice artifacts can be significant. Therefore, it is important to remove
these artifacts and to estimate the corresponding systematic errors before
comparing the lattice results with perturbation theory.
Since we are interested in the cutoff effects at short distances, we could use
lattice perturbation theory to estimate these effects, due to asymptotic freedom in QCD. 
In the simplest case we could use the tree-level result.
At tree level the static energy on the lattice is given by
\begin{eqnarray}
&
\displaystyle
E_0^{lat,tree}(r)=-C_F g^2 \int\frac{d^3 \mathbf{k}}{(2 \pi)^3} D_{00}(k_0=0,\mathbf{k}) e^{i \mathbf{k} \mathbf{r}}\nonumber\\
&
\displaystyle
=:-C_F g^2 \frac{1}{4 \pi r_I(r)},
\label{tree}
\end{eqnarray}
where $D_{00}(k_0=0,\mathbf{k})$ is the temporal gluon propagator on the lattice, 
\begin{equation}\label{eq:D00latt}
D_{00}^{-1}(k_0=0,\mathbf{k})=4 \sum_{i=1}^3 \left( \sin^2\frac{k_i}{2} + c_{w} \sin^4\frac{k_i}{2} \right),
\end{equation}
which
depends on the choice of the gauge action, and $C_F$ is the Casimir of the fundamental representation. Equation~(\ref{eq:D00latt}) is given in lattice-spacing units. For unimpoved gauge action $c_w=0$, while for the improved gauge action
$c_w=1/3$ \cite{Weisz:1982zw,Weisz:1983bn}. The integral in Eq.~(\ref{tree}) has to 
be evaluated numerically. Equation~(\ref{tree}) defines the so-called improved distances $r_I(r)$ (for simplicity, from now on we omit the dependence on $r$ and just write $r_I$), that renders the lattice Coulomb potential
to be continuum-like and will be used below to reduce lattice artifacts.
In Fig. \ref{fig:free} we show the ratio of the lattice static energy to the continuum static energy at tree level. As one can see, the discretization effects are small, at
the level of few percent, even at distance $r/a=1$. For the improved gauge action the discretization
effects are below 1\% level for $r/a>2$, while they are still significant for the standard gauge action.
To estimate the cutoff effects in the actual lattice calculations we need a continuum
estimate of the static energy. Based on the free-theory
result we assume that the cutoff effects are negligible for distances $r/a>2$, and fit the
lattice results at $\beta=7.825$ in this region 
to Coulomb plus linear plus constant form, to obtain a continuum estimate. 
To reduce the cutoff effects we replace the distance $r$ by $r_I$ in the fit (see below).
The fit can be used as a continuum estimate for distances $r>0.3r_1$.
In order to compare results
at different $\beta$ values to the continuum estimate, we normalize the static energy in units of the $r_1$ scale using
the normalization procedure described in Ref.~\cite{Bazavov:2012ka}. 
\begin{figure*}
\includegraphics[width=8.6cm]{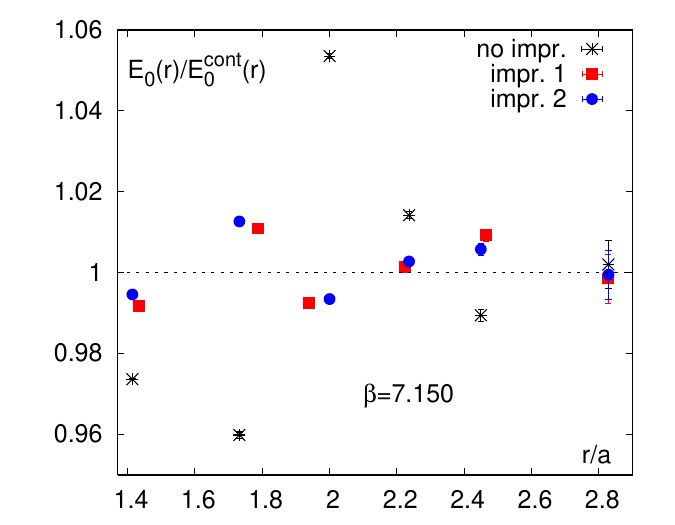}
\includegraphics[width=8.6cm]{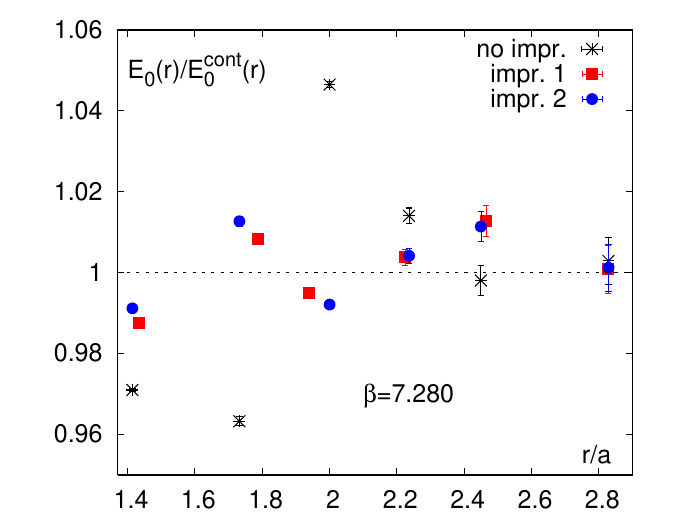}
\includegraphics[width=8.6cm]{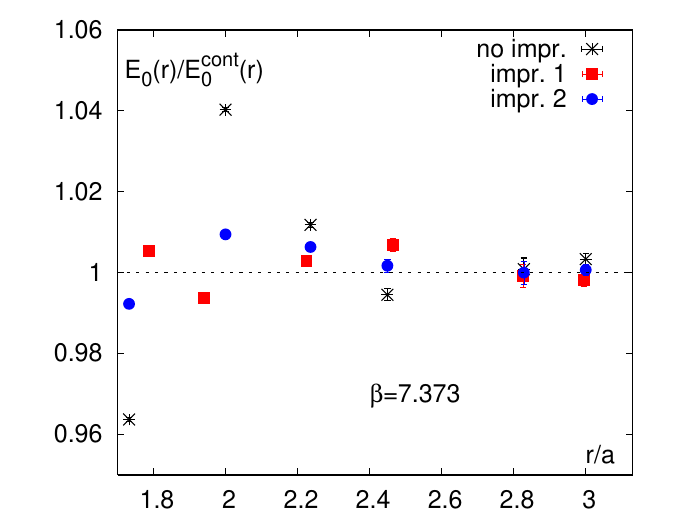}
\includegraphics[width=8.6cm]{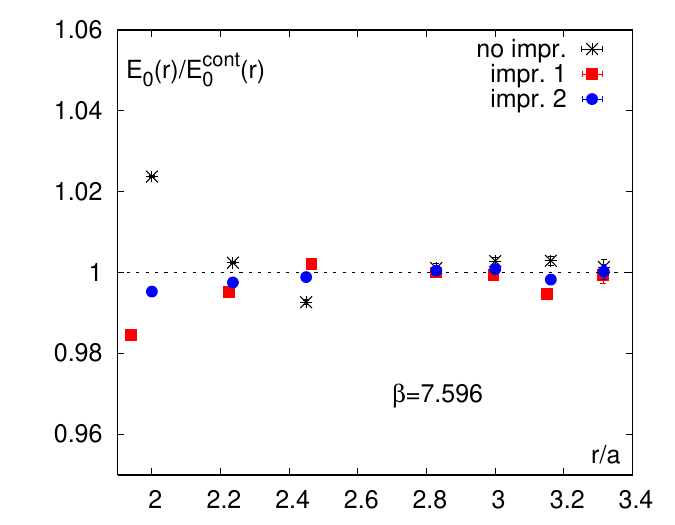}
\caption{The ratio of the static energy at $\beta=7.150,~7.280,~7.373,~7.596$ to the continuum estimate.
The bursts correspond to the un-corrected static energy. }
\label{fig:rat}
\end{figure*}
In Fig.~\ref{fig:rat} we show the ratio of the static energy at different lattice spacings ($\beta$)
to the continuum estimate described above for $r>0.3r_1$. In terms of $r$ 
in lattice units we cover the distance range $r/a>\sqrt{2}$. Obviously, the range in $r/a$ extends
to smaller values for smaller values of $\beta$.
As one can see, the discretization effects
are similar to the free theory, although slightly larger in some cases. Thus, the tree-level result gives
us a fair estimate on the size of discretization effects, as expected. We can use the free-theory
result to correct for the lattice artifacts at short distances, i.e. implement a tree-level improvement of the
static energy. One way to correct for the lattice
artifacts is to replace the distance $r$ by the improved distance $r_I$ defined in Eq.~(\ref{tree}).
We will refer to this improvement scheme as improvement 1. Alternatively, we can fit the static energy to the form
\begin{equation}
V(r)=-\frac{a}{r}+\sigma r+C+a' \left(\frac{1}{r}-\frac{1}{r_I}\right),
\end{equation}  
and correct the lattice data by subtracting $a' (1/r-1/r_I)$. We call this improvement scheme
improvement 2. 
We see from Fig.~\ref{fig:rat} that the tree-level corrections reduce the discretization errors to the level of 1.5\% or smaller. Furthermore,
the deviations from the continuum estimate are roughly independent of $\beta$ in the considered range of the gauge couplings. From the figure, one can also see that the two improvement schemes give similar results.
We adopt  improvement scheme 1 and try to correct for the residual cutoff effects by dividing
the value of the static energy calculated on the lattice with the correction factors given
in Table~\ref{tab:corr}.
\begin{table*}
\begin{tabular}{|c|cccccc|}
\hline
$\beta$ &  $r/a=1$    &   $r/a=\sqrt{2}$  & $r/a=\sqrt{3}$ & $r/a=2$  & $r/a=\sqrt{5}$  & $r/a=\sqrt{6}$  \\
\hline
7.150   &  0.980      &   0.995           & 1.007          & 0.988    & 1.000           & 1.010           \\
7.280   &  0.980      &   0.997           & 1.008          & 0.992    & 1.000           & 1.013           \\
7.373   &  0.980      &   0.998           & 1.009          & 0.994    & 0.995           & 1.005           \\
7.596   &  0.980      &   0.995           & 1.005          & 0.994    & 1.000           & 1.001           \\
7.825   &  0.968      &   0.992           & 1.005          & 0.994    & 0.998           & 1.001           \\
\hline
\end{tabular}
\caption{The correction factors for the static energy for the six smallest
lattice distances.}
\label{tab:corr}
\end{table*}
The correction factors have been estimated using an iterative procedure as follows. First we calculate
the ratios of the static energy to the above continuum estimate for $r/a=\sqrt{2}$ and $\beta=7.280$,
$r/a=\sqrt{3}$ and $\beta=7.373$, and $r/a=2$ and $\beta=7.596$, and divide the lattice results
on the static energy for $\beta=7.825$ for $r/a=\sqrt{2},~\sqrt{3}$ and $2$ by the corresponding ratios,
i.e. we use the lattice data at the closest value of $\beta$ to estimate the correction factor.
Furthermore, we assign a systematic error of $0.5\%$ to these data points and fit the corrected data
points for $\beta=7.825$ by a Coulomb plus linear plus constant form to obtain the improved
continuum estimate of the static energy, which now extends to smaller distances, namely down to
distance $r=0.186r_1$. With this continuum estimate we calculate the correction factor for the smallest
distance $r/a=1$ by dividing the $\beta=7.373$ data at $r/a=1$ by the corresponding value. Using
this correction factor for the $\beta=7.825$ static energy at $r/a=1$, and assigning a systematic
error of $1\%$ to that point, we could extend the fit to
the smallest lattice distance to obtain our final continuum estimate for the static energy for
distances $0.12< r/r_1 < 0.6$ (note that the upper limit of the fit range was kept fixed in all the iterative steps), and also obtain the correction factors given in Table \ref{tab:corr}.
We assign a systematic error of $1\%$ for all the corrected lattice data at distance $r/a=1$ and
a systematic error of $0.5\%$ for all the remaining 5 shortest distances. The corrected lattice
data for the static energy with the statistical and systematic errors added in quadrature is
shown in Fig.~\ref{fig:ratall}. The figure clearly shows that, within estimated errors, all
the lattice data on the static energy agree with the continuum estimate, i.e. our final
error estimates are sufficiently large to included all possible residual discretization effects.

\begin{figure}
\includegraphics[width=8.6cm]{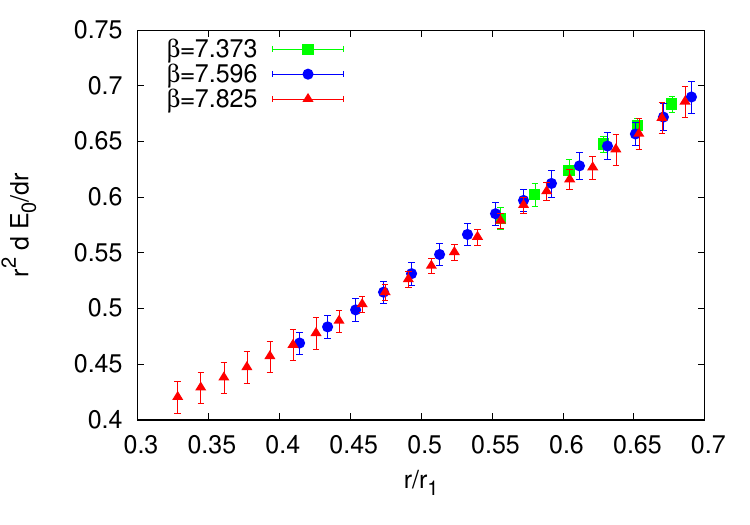}
\caption{The force calculated on the lattice for $\beta=7.373,~7.596$ and $7.825$.}
\label{fig:force}
\end{figure}

We have also calculated the force from the lattice data at the three smallest lattice spacings. The calculation of the derivative of the static energy from a discrete set of lattice data is non-trivial, since both discretization errors and statistical fluctuations result in a non-smooth behavior of the static energy as function of the distance $r$, and subsequently in non-monotonic behavior of the force. To avoid problems with the lattice artifacts, we calculate the force only for distances $r/a>2$, for which discretization effects are expected to be small. To obtain a result for the force from the lattice data on the static energy which is monotonic, we fit the latter with smoothing splines. The smoothing spline is determined by minimizing the $\chi^2$ plus the integral of the second derivative
of the fit function in the considered interval times a real parameter $\lambda$.
The spline fit is performed using the R statistical package \cite{Rpackage}.
The errors on the spline are calculated using standard bootstrap method, i.e. for each data
point on the static energy we generate a set of synthetic data according to a Gaussian 
distribution, with the width and mean given by the statistical error and central value 
of the lattice data, perform the spline fit on the synthetic data, and finally
perform the statistical analysis of the resulting splines.
Furthermore,  we consider several
values of the smoothing parameter $\lambda$ in the standard range $0<\lambda<1$. We find
that, within statistical errors, there is little dependence on the smoothing parameter.
For the distances where the dependence on $\lambda$ was larger than the statistical errors,
the final error estimate was increased to accommodate the difference between different splines.
In Fig.~\ref{fig:force} we show the lattice results on the force for $\beta=7.373,~7.596$ and $7.825$
from the spline fit evaluated at distances separated by $a/8$ in the interval $0.32 < r/r_1 < 0.7$, with the corresponding statistical errors.
As one can see from the figure, the values for the force obtained for different $\beta$ agree well within 
the estimated errors, confirming our expectation that cutoff effects are small for $r/a>2$. The data points
for the force shown in Fig.~\ref{fig:force} will be used in the analysis presented in Sec.~\ref{subsubsec:compforce}.

\section{Perturbative expressions}\label{sec:pert}
In this section we detail the perturbative expressions that are used in our analyses.

Let us start by recalling that our goal is to compare with the lattice data for the static energy. An important point to remember, to that effect, is that the normalization of the static energy computed on the lattice is not physical, but only its slope is. At the same time, care needs to be taken with the so-called renormalon singularities, present in the perturbative expressions. At the end of the day, the renormalon has no practical effect in the extraction, i.e. the comparisons, and any fits to the lattice data involved in the analyses, can always be done in a way such that they are not affected by the presence of the renormalon. This is again due to the fact that the overall normalization of the energy is affected by the renormalon, but the slope, which is what matters for the extraction of $\alpha_s$, is renormalon free. It is nevertheless convenient to keep track of the presence of the renormalon in intermediate expressions, and to always work with a scheme where the renormalon is subtracted explicitly. We can schematically write the static energy as
\begin{equation}\label{eq:E0sch}
E_0(r)=V_s(r,\nu,\mu)+\delta_{US}(r,\nu,\mu)+RS(\rho)+\Lambda_s(\rho),
\end{equation}
where $V_s(r,\nu,\mu)$ is the static potential, $\delta_{US}(r,\nu,\mu)$ represents the ultrasoft contributions to the energy (which first appear at three-loop order, see Eq.~(\ref{eq:pertF}) below), $RS(\rho)$ generically represents a term that cancels the leading renormalon singularity in $V_s(r,\nu,\mu)$, and $\Lambda_s(\rho)$ contains the residual mass term that is necessary to define the static limit of QCD. The $\nu$ dependence in $V_s$ and $\delta_{US}$ cancels order by order in perturbation theory. We mention that $\mu$ is the ultrasoft factorization scale, but it does not play a relevant role for the discussion in this paragraph. For concreteness, we adopt the Renormalon Subtracted (RS) scheme~\cite{Pineda:2002se}. The expression for $V_s(r,\nu,\mu)$ is given as a series in $\als(\nu)$ and contains $\ln(r\nu)$ terms, starting at order $\als^2(\nu)\ln(r\nu)$. The expression for $RS(\rho)$ is given as a series in $\als(\rho)$. In order to cancel the renormalon order by order, $\als(\rho)$ needs to be expanded in terms of $\als(\nu)$ (or vice versa); when one does that, $RS(\rho)$ is given as well as a series in $\als(\nu)$ and contains $\ln(\rho/\nu)$ terms, starting at order $\als^3(\nu)\ln(\rho/\nu)$. If one sets $\nu=1/r$ in Eq.~(\ref{eq:E0sch}), one resums the $\ln(r\nu)$ logarithms in $V_s(r,\nu,\mu)$, but on the other hand the RS-subtraction term is not constant. Since the normalization of the renormalon singularity, which enters in $RS(\rho)$, can only be computed approximately, that would introduce additional uncertainties in the perturbative expression, that would hinder any precise $\alpha_s$ extraction. One can also set $\nu$ constant in Eq.~(\ref{eq:E0sch}), in that case the RS-subtraction term is itself constant, and gets absorbed in the additive constant needed to compare with lattice data (see Eq.~(\ref{eq:Elattcomp}) below), but one has $\ln(r\nu)$ terms left in $V_s(r,\nu,\mu)$. If one only compares with lattice data in a restricted distance range, one is free to set $\nu$ at the center of that distance range and the logarithms never become large. This last option is the approach that was taken in Ref.~\cite{Bazavov:2012ka}. Alternatively, one can also take a derivative of Eq.~(\ref{eq:E0sch}), and compute the force. We can take the derivative before resumming the $\ln(r\nu)$ terms, i.e. from Eq.~(\ref{eq:E0sch}) with constant $\nu$, so that the RS-subtraction term, and $\Lambda_s$, completely disappear. That is the procedure we follow here.

By taking a derivative, with respect to $r$, of Eq.~(\ref{eq:E0sch}) we obtain the force,
\begin{equation}\label{eq:forcenu}
F(r,\nu)=\frac{dE_0}{dr}=\frac{C_F}{r^2}\left[\alpha_E(r,\nu)-r\alpha'_E(r,\nu)\right],
\end{equation}
where $\alpha'_E(r,\nu)$ denotes the derivative of $\alpha_E(r,\nu)$ with respect to
$r$, and where we defined
\begin{equation}
V_s(r,\nu,\mu)+\delta_{US}(r,\nu,\mu)=:-\frac{C_F}{r}\alpha_E(r,\nu).
\end{equation}
At this point the expression for the force $F(r,\nu)$ is given as a series in $\als(\nu)$ and contains $\ln(r\nu)$ terms, starting at order $\als^2(\nu)\ln(r\nu)$. We can now resum these logarithms by setting $\nu=1/r$, i.e. we re-organize the perturbative expansion for the force as a series in $\als(1/r)$. Having obtained this expression, one can either directly compare $F(r,\nu=1/r)$ with lattice data for the force, or numerically integrate $F(r,\nu=1/r)$ and compare with lattice data for the energy, that is, to numerically compute
\begin{equation}\label{eq:EfrF}
E_0(r)=\int_{r_*}^{r}F(r,\nu=1/r)\, dr,
\end{equation}
where the lower limit of integration, $r_*$, is irrelevant because
the constant contribution coming from it gets absorbed in the additive
constant used to compare with lattice data; i.e. when we compare with lattice data we need to plot
\begin{equation}\label{eq:Elattcomp}
E_0(r)-E_0(r_{\rm ref})+E_0^{\rm latt.}(r_{\rm
  ref})=E_0(r)+\textrm{const.},
\end{equation}
where $r_{\rm ref}$ is the reference distance where we make the perturbative expression coincide with the lattice data, and $E_0^{\rm latt.}(r_{\rm ref})$ is the value of the static energy computed on the lattice at that distance. As it will be further explained in Sec.~\ref{sec:asextr}, we will use the comparison with the lattice data for the energy, through Eqs.~(\ref{eq:EfrF})-(\ref{eq:Elattcomp}), for our main analyses, and leave the direct comparison of $F(r,\nu=1/r)$ with the lattice data for the force as a cross check of our results.

The perturbative expansion of $F(r,\nu=1/r)$ is explicitly given by
\begin{eqnarray}\label{eq:pertF}
F(r,\frac{1}{r}) & = & \frac{C_F}{r^2}\als(1/r)\Bigg[1\nonumber\\
&&+\frac{\als(1/r)}{4\pi}\Big(\tilde{a}_1-2\beta_0\Big)\nonumber\\
&&+\frac{\als^2(1/r)}{(4\pi)^2}\Big(\tilde{a}_2-4\tilde{a}_1\beta_0-2\beta_1\Big)\nonumber\\
  &&+\frac{\als^3(1/r)}{(4\pi)^3}\Big(\tilde{a}_3-6\tilde{a}_2\beta_0-4\tilde{a}_1\beta_1-2\beta_2\nonumber\\
&&+a_3^L\ln\frac{C_A\als(1/r)}{2}\Big)+\mathcal{O}(\als^4,\als^4\ln^2\als)\Bigg],
\end{eqnarray}
where tree-level accuracy corresponds to taking the first line of the equation, and $i$-loop order to taking terms up to order $\als^i$ in the square brackets. The coefficients $\tilde{a}_i$ in the equation above have been calculated over the years, and are currently known up to three-loop order, i.e. $i=3$~\cite{Fischler:1977yf,Billoire:1979ih,Peter:1996ig,Peter:1997me,Schroder:1998vy,Kniehl:2001ju,Smirnov:2008pn,Anzai:2009tm,Smirnov:2009fh}. The existence of the $\ln\als$ terms was noticed a long time ago~\cite{Appelquist:1977es}, and the $a_3^L$ coefficient was computed in Refs.~\cite{Brambilla:1999qa,Kniehl:1999ud}; the corresponding coefficient at the next order, $a_4^L$, is also known~\cite{Brambilla:2006wp}. These $\ln\als$ terms, so-called ultrasoft logarithms, have also been resummed, to sub-leading accuracy; we present the ultrasoft-resummed expressions in Sec.~\ref{subsec:resus}. The values of the $\tilde{a}_i$, $a_i^L$ coefficients have been collected in Ref.~\cite{Tormo:2013tha}, and we will not copy the expressions here. We refer to that reference, which uses exactly the same notation we employ here, for the concrete expressions; $\beta_i$ are the coefficients of the beta function and their expressions, as well as those of the color factors, are also collected there.

As mentioned above, we will use Eqs.~(\ref{eq:EfrF})-(\ref{eq:pertF}) for our analyses. Note again that all reference to the renormalon is gone in those expressions. These are essentially the same expressions that are used in related analyses by, for instance, the ALPHA collaboration~\cite{Leder:2011pz,Donnellan:2010mx}, where no explicit mention to renormalons is made at all. Keeping track of the presence of the renormalon is just a convenient way to identify which quantities have a well-behaved perturbative expansion, and are adequate for a precise $\als$ extraction. Recall that, in our previous analysis~\cite{Bazavov:2012ka} we used the perturbative expression coming from Eq.~(\ref{eq:E0sch}) with $\nu$ constant. In that case the renormalon got absorbed in the additive constant that is added to $E_0(r)$ to compare with lattice, and did not affect the fits either. The advantage of the method pursued in the present analysis, with respect to the one in Ref.~\cite{Bazavov:2012ka}, is that $\ln(r\nu)$ terms are completely absent. When one compares with lattice data in a restricted $r$ range, one can always choose $\nu$ at the middle of the range, and the logs do not become large, but since in the present analysis we reach shorter distances than in Ref.~\cite{Bazavov:2012ka}, it is better suited to use an expression that completely avoids all $\ln(r\nu)$ terms, as we do now. All these different options, in treating the perturbative expressions when comparing with lattice data, were already explained in Ref.~\cite{Pineda:2002se}; the method we use here, through Eqs.~(\ref{eq:EfrF})-(\ref{eq:pertF}), was mentioned but not really employed, though, presumably because the aim of that paper was mostly to study the presence of the renormalon, rather than a precise determination of a parameter.

We also mention that we explicitly checked that finite strange-quark mass effects can be neglected when comparing with lattice data, therefore we will safely ignore them in the present analysis. Finally we note that we use four-loop accuracy for the running of $\als$ everywhere.

\subsection{Resummation of the ultrasoft logarithms}\label{subsec:resus}
The ultrasoft logarithms, i.e. the $\ln\als$ terms, in the static energy, were resummed at leading order in Ref.~\cite{Pineda:2000gza} and at sub-leading order in Ref.~\cite{Brambilla:2009bi}. We also employ the ultrasoft-resummed expressions in our analyses here. 

These ultrasoft terms can be conveniently resummed by solving Renormalization Group (RG) equations in the effective theory potential Non-Relativistic QCD (pNRQCD)~\cite{Brambilla:1999xf,Brambilla:2004jw}. The RG equations in pNRQCD have been written at the level of the matching coefficients, i.e. $V_s$ and $\Lambda_s$ in Eq.~(\ref{eq:E0sch}). Therefore, to obtain the expressions we need to compare with lattice data, what we do is the following: We first resum the ultrasoft logarithms for the potential, i.e. $V_s$, and $\Lambda_s$, by solving the RG equations in pNRQCD from an initial scale $\mu_0=\nu$, to a scale $\mu$. We then take the derivative of the static energy with the ultrasoft-resummed expressions to obtain the force. Finally, we re-organize the perturbative expansion for the force as a series in $\als(1/r)$. 

After doing that, the perturbative expression for the force is given by\footnote{We thank Antonio Pineda for making us aware of a misprint in earlier versions of this formula~\cite{Ayala:2020odx}.}

\begin{eqnarray}\label{eq:pertFus}
F(r,\frac{1}{r}) & = & \frac{C_F}{r^2}\als(1/r)\Bigg[1\nonumber\\
&&+\frac{\als(1/r)}{4\pi}\Big(\tilde{a}_1-2\beta_0\Big)\nonumber\\
&&+\frac{\als^2(1/r)}{(4\pi)^2}\Big(\tilde{a}_2-4\tilde{a}_1\beta_0-2\beta_1\Big)\nonumber\\
  &&+\frac{\als^3(1/r)}{(4\pi)^3}\Big(\tilde{a}_3-6\tilde{a}_2\beta_0-4\tilde{a}_1\beta_1-2\beta_2\Big)\nonumber\\
&& -\frac{\als^2(1/r)}{(4\pi)^2}\frac{a_3^L}{2\beta_0}\ln\frac{\als(\mu)}{\als(1/r)}\nonumber\\
  && -\frac{\als^2(1/r)}{(4\pi)^2}\frac{4C_A^3\pi^2}{3}\left(\eta_0-\frac{1}{\pi}\left(\frac{1}{6}+\ln2\right)\right)\nonumber\\
&&  \times\left(\als(\mu)-\als(1/r)\right)\nonumber\\
&& +\frac{\als^3(1/r)}{(4\pi)^3}8C_A^3\pi^2\left(2-\frac{\tilde{a}_1}{\beta_0}\right)\ln\frac{\als(\mu)}{\als(1/r)}\nonumber\\
&& +\frac{\als^2(1/r)\als(\mu)}{(4\pi)^3}a_3^L\ln\frac{C_A\als(1/r)}{2r\mu}\nonumber\\
&& +\mathcal{O}(\als^4,\als^5\ln\als)\Bigg],
\end{eqnarray}
where we refer again to Ref.~\cite{Tormo:2013tha} for a concrete expression of the coefficient $\eta_0$. The terms that are displayed in Eq.~(\ref{eq:pertFus}) account for all the contributions that are needed up to order $\alpha_s^{4+n}\log^n\alpha_s$ ($n\ge0$), which is what we call next-to-next-to-next-to-leading logarithmic
(N$^3$LL) accuracy; accordingly, next-to-next-to-leading logarithmic (N$^2$LL) accuracy includes the contributions up to order $\alpha_s^{3+n}\log^n\alpha_s$, and corresponds to the first, second, third, and fifth lines in Eq.~(\ref{eq:pertFus}). Note that the, scheme-dependent, constant $K_2$ that was present at N$^3$LL accuracy in the analysis of Ref.~\cite{Bazavov:2012ka} is absent in the scheme we use in the present analysis. Recall that $\mu$ is the ultrasoft factorization scale, which takes a natural value $\mu\sim(C_A\als)/(2r)$, and that the apparent $\mu$ dependence in Eq.~(\ref{eq:pertFus}) is of higher order in $\alpha_s$.

\section{Analyses for {\large $\alpha_s$} extraction}\label{sec:asextr}
Our aim is to compare the lattice data with the perturbative expressions, as presented in the previous sections, with the goal of obtaining a determination of $\als$. The general concept of performing such an analysis was suggested long ago~\cite{Michael:1992nj,Booth:1992bm}, but it is only the recent progress in the evaluation of the static energy, as summarized in the previous sections, that allowed for a practical realization~\cite{Bazavov:2012ka}. The basic idea we follow here is very simple: the perturbative expressions depend on $\Lambda_{\MS}$, where $\Lambda_{\MS}$ is the QCD scale (which fixes $\als$), and we can use the comparison to find the values of $r_1\Lambda_{\MS}$ that are allowed by lattice data. Here, $r_1$ is the reference scale used in the lattice computation, its value has also been independently computed on the lattice, and using it we obtain $\Lambda_{\MS}$ in physical units. To find the values of $r_1\Lambda_{\MS}$ allowed by lattice data, one goes to the short-distance region, where it is expected that perturbation theory is enough to describe the lattice results. The guiding principle we follow, to find the allowed $\als$ values, is that the agreement of the theoretical predictions with lattice data should improve when the perturbative order of the computation is increased. At the same time, the error that one assigns to the result should be such that it reflects the uncertainties due to unknown higher perturbative orders.

As explained in the previous sections, to perform such an analysis, it is important to remember that the normalization of the lattice result is not physical, but only the slope is. Therefore one must either normalize the energy to a certain value at a given distance, as specified by Eq.~(\ref{eq:Elattcomp}), or compare directly with the force. Additionally, since we have lattice results for several different lattice spacings, see Sec.~\ref{sec:lattice}, one can either put all the lattice data together, or perform an analysis for each lattice spacing separately and then take the average. Our previous analysis in Ref.~\cite{Bazavov:2012ka} employed the procedure devised in Ref.~\cite{Brambilla:2010pp} to implement the guiding principle of improved agreement with increasing perturbative orders, used all lattice data together, and compared with lattice according to Eq.~(\ref{eq:Elattcomp}).

One of the crucial aspects in this kind of analyses is always to know whether the current lattice data has really reached the purely perturbative regime, and with enough precision to perform the extraction, or not. It is not an easy task to undoubtedly state this point. In that sense, the facts that the analyses of Ref.~\cite{Bazavov:2012ka} showed that the agreement with lattice indeed improved when increasing the perturbative order, and that the perturbative curves were able to describe the lattice data quite well, can be seen as positive evidence that this was the case. Thanks to the enlarged lattice data set that we have in the present paper, which reaches shorter distances, we can now perform a more detailed analysis of this issue.

\subsection{Procedure to extract $\als$}\label{subsec:procextr}
With the main motivation of further testing if the lattice data has reached the purely perturbative regime, we modify the procedure we used in Ref.~\cite{Bazavov:2012ka} for the extraction of $\als$, and proceed as we describe next. 

First of all, we use now the data for each value of the lattice spacing separately, and at the end perform an average of the different obtained values of $\als$. The reason for proceeding this way is that, when one wants to put all lattice data together, one needs to normalize the results calculated at different lattice spacings to a common value at a certain distance, due to the additive ultraviolet renormalization of the static energy. There is some uncertainty related to this normalization procedure, and in fact, the errors on the lattice data due to the normalization are larger than the lattice systematic errors. On the other hand, if one uses the data for each lattice spacing separately, there is no need to normalize it, and the data errors are, therefore, smaller. The downside of this option is that, of course, one has less data in each analysis. With the enlarged data set we use in this paper, we have enough data at short distances to perform analyses for each lattice spacing separately. This was not quite the case with the data set used in Ref.~\cite{Bazavov:2012ka}.

\begin{figure}
\centering
\includegraphics[width=8.6cm]{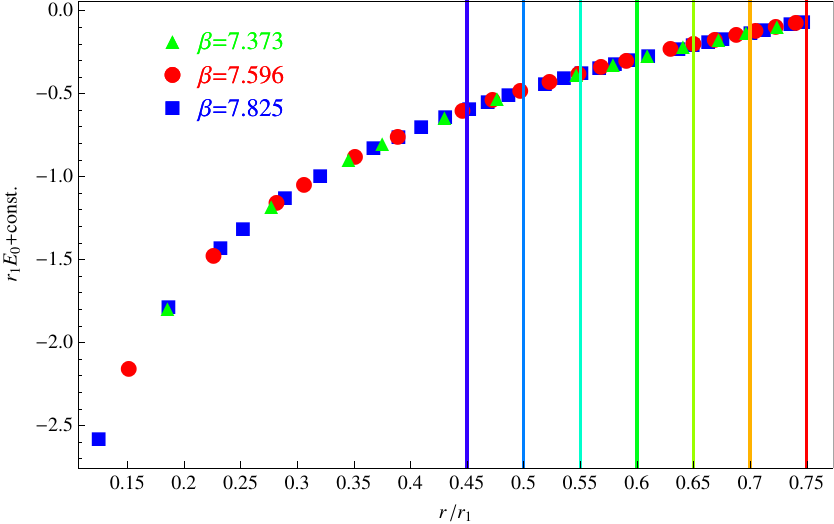}
\caption{Data sets used in the analyses, with the different distance ranges that we use in the fits marked as the vertical lines (ranges are from each of the vertical lines to the left).}\label{fig:dataranges}
\end{figure}

The procedure used in Ref.~\cite{Bazavov:2012ka} was devised to provide a faithful estimate of the central value and error of the result~\cite{Brambilla:2010pp}. Here we modify it to be able to further test if we have reached the purely perturbative regime, while maintaining the reliability of the outcome. At the same time, we try to keep the whole procedure as simple and straightforward as possible. In Ref.~\cite{Bazavov:2012ka} we used the distance region $r<0.75r_1$ for all our analyses and fits. We now use $r<0.75r_1$ as the largest distance range we reach, but also perform analyses with points at shorter distance ranges only. We proceed as follows:
\begin{enumerate}
\item Perform fits to the lattice data for the static energy, according to Eqs.~(\ref{eq:EfrF})-(\ref{eq:Elattcomp}), at different orders of perturbative accuracy. The parameter of the fits is $\Lambda_{\MS}$.
\item Repeat the above fits for each of the following distance ranges: $r<0.75r_1$, $r<0.7r_1$, $r<0.65r_1$, $r<0.6r_1$, $r<0.55r_1$, $r<0.5r_1$, and $r<0.45r_1$.
\item Use the ranges where the reduced $\chi^2$ of the fits either decreases (or remains almost constant, i.e. it does not increase by more than one unit) when increasing the perturbative order, or is smaller than~1.
\item To estimate the perturbative uncertainty of the result, we repeat the fits in two ways: (i) by varying the scale in the perturbative expansion, from $\nu=1/r$ to $\nu=\sqrt{2}/r$ and $\nu=1/(\sqrt{2}r)$, and (ii) by adding/subtracting a term $\pm(C_F/r^2)\als^{n+2}$ to the expression at $n$ loops.
\end{enumerate}
As mentioned, we use the data for each lattice spacing separately. This means that, in order to have a significant number of data points for each of the distance ranges above, we only use the $\beta=7.373, 7.596, 7.825$ sets. To illustrate how many points enter in each range, we show these data sets, with the distance ranges marked, in Fig.~\ref{fig:dataranges}; note that this figure just aims at showing how many points we have, and, therefore, everything is displayed in units of the scale $r_1$ and the different data sets are normalized, but no error associated to that is shown.
The actual fits are performed with the data for each set in units of the corresponding lattice spacing, $a$, and then the result is translated to units of $r_1$ using the corresponding value of $r_1/a$ and its error, as given in Tab.~\ref{tab:r1}. We always use fits that consider the lattice data for a given $\beta$ to be uncorrelated. This is sufficient since we mostly consider the static energy at off-axis values of $r$, for which the off-diagonal elements of the correlation matrix are small. Note that the two sets of fits in point 4 of the list above both provide an estimate of the uncertainty due to unknown higher-order perturbative terms; we will take whichever of the two gives a larger uncertainty as the perturbative-error estimate. In addition, the fits using different distance ranges probe $\als$ at different scales, and agreement among their results would be an indication that perturbative uncertainties are properly estimated.

\subsection{Outcomes of the analyses}\label{subsec:outana}
We present in this section the outcomes of the fits described above. We compare with lattice data for the static energy according to Eq.~(\ref{eq:Elattcomp}). For that we need to choose which lattice point defines $r_{\rm ref}$. In principle one would choose $r_{\rm ref}$ as the shortest distance where lattice data is available, since the perturbative expressions should be more reliable there. In practice though, as discussed in detail in Sec.~\ref{sec:lattice}, the lattice points at shorter distances have larger systematic discretization errors. Since when we normalize the perturbative curves to $r_{\rm ref}$ we effectively put the error of that lattice point to zero, it would not be appropriate to use one of the points with larger systematic errors to define $r_{\rm ref}$. Therefore, we do not use any of the first six points at each lattice spacing to define $r_{\rm ref}$. Also, to make sure that the results of the fits are not sensitive to the point we choose to define $r_{\rm ref}$, one should perform analyses with different choices of $r_{\rm ref}$. According to these discussions, we choose the seventh, eighth, or ninth point at each lattice spacing to define $r_{\rm ref}$; if for some data set at some distance range there are not enough points to do that, we choose the point preceding it in distance instead, i.e. if, for example, in a given distance range there were only six points we would choose the sixth point to define $r_{\rm ref}$.

\begin{figure}
\centering
\includegraphics[width=8.6cm]{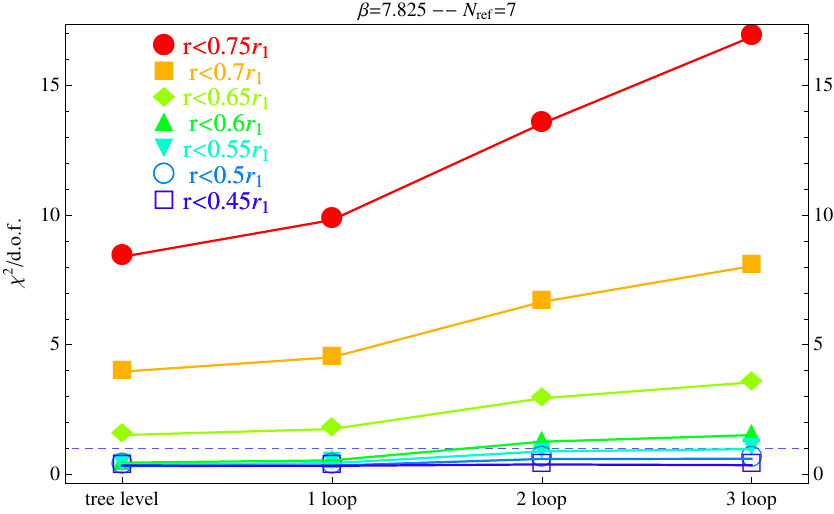}
\includegraphics[width=8.6cm]{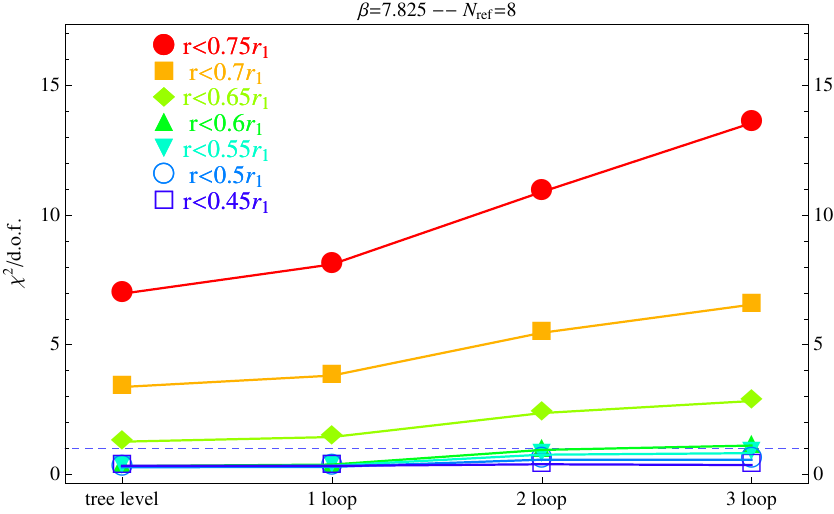}
\includegraphics[width=8.6cm]{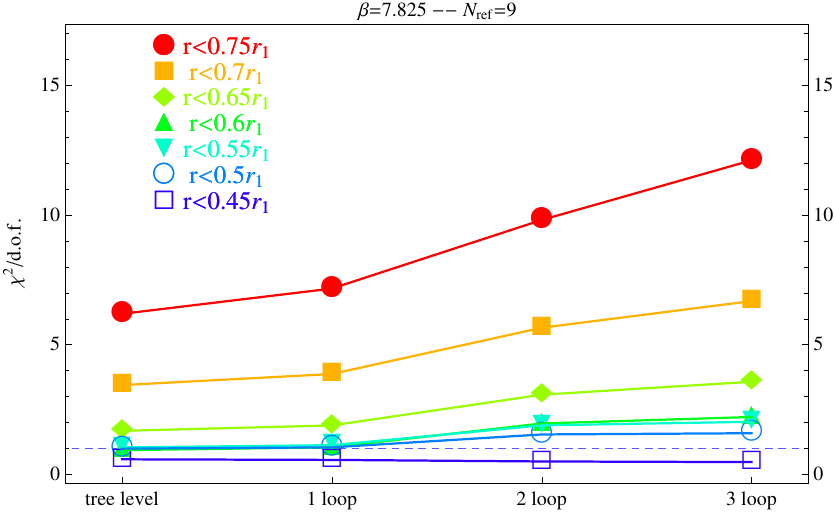}
\caption{Reduced $\chi^2$ of the fits at different orders of perturbative accuracy, and for several distance ranges. The dashed blue line marks $\chi^2/{\rm d.o.f.}=1$ for reference. All three panels correspond to the $\beta=7.825$ data set. $N_{\rm ref}$ is the lattice point that defines $r_{\rm ref}$ in Eq.~(\ref{eq:Elattcomp}), counting from the point at shortest distance.}\label{fig:chisfits_beta7825}
\end{figure}

\begin{figure}
\centering
\includegraphics[width=8.6cm]{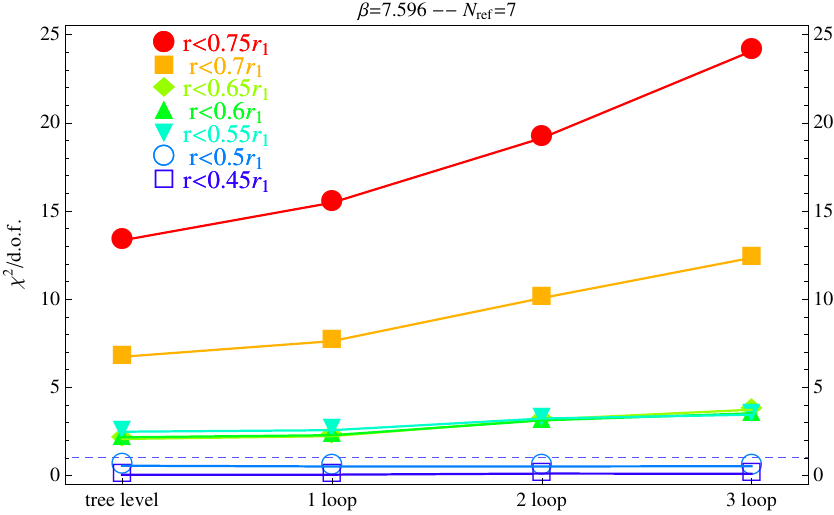}
\includegraphics[width=8.6cm]{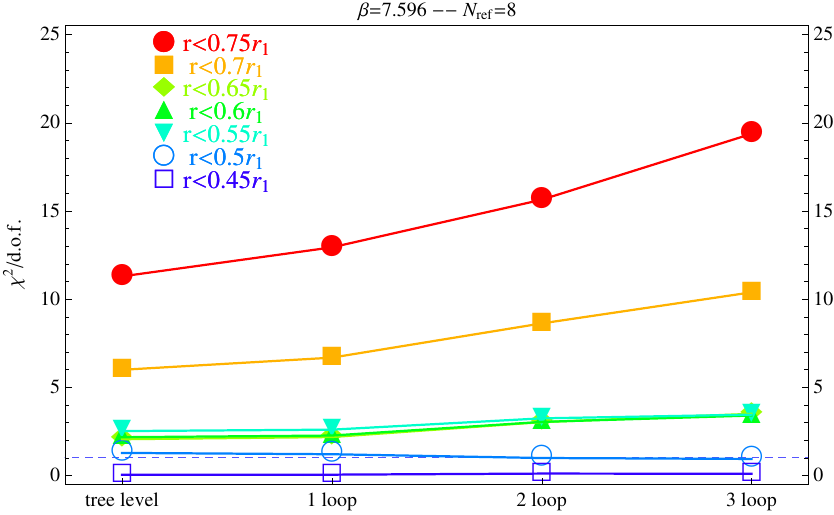}
\includegraphics[width=8.6cm]{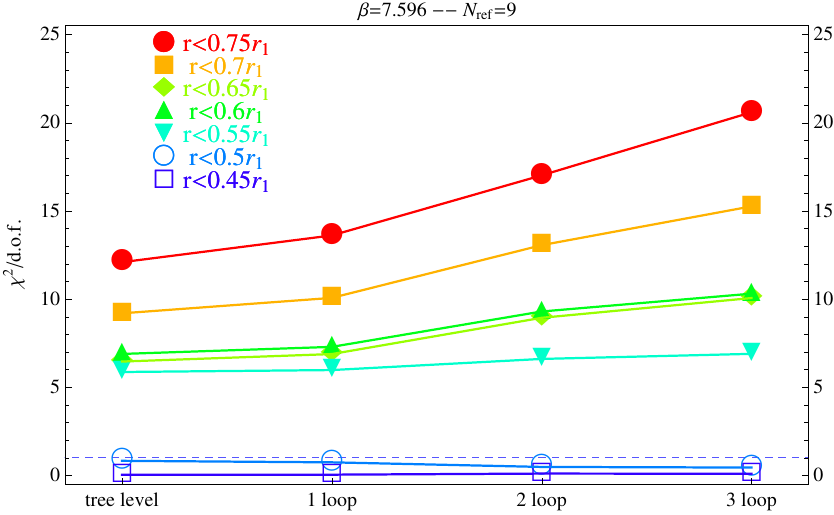}
\caption{Same as Fig.~\ref{fig:chisfits_beta7825} for $\beta=7.596$.}\label{fig:chisfits_beta7596}
\end{figure}

\begin{figure}
\centering
\includegraphics[width=8.6cm]{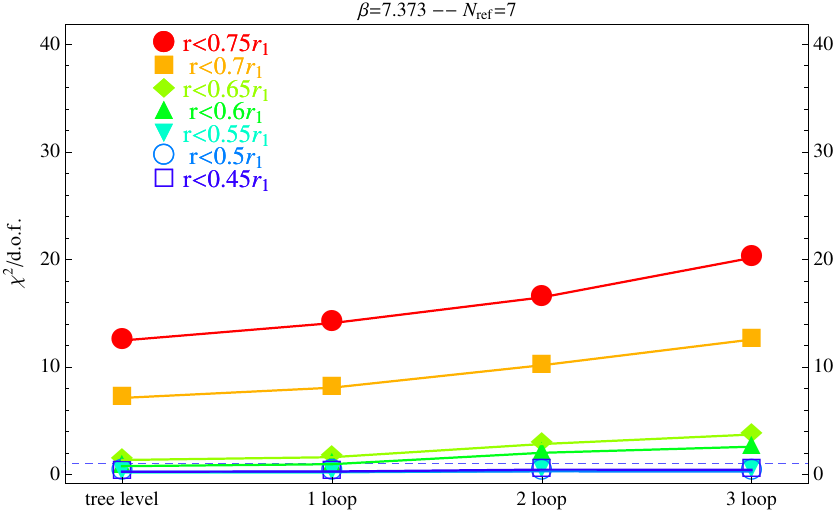}
\includegraphics[width=8.6cm]{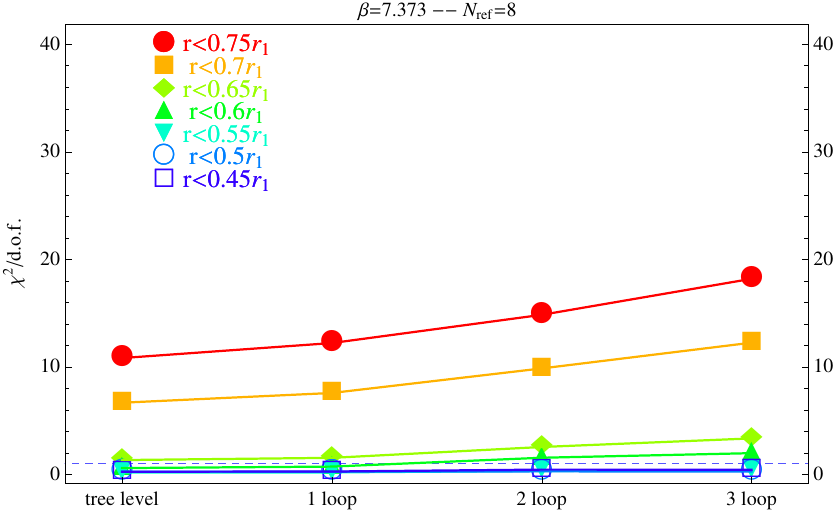}
\includegraphics[width=8.6cm]{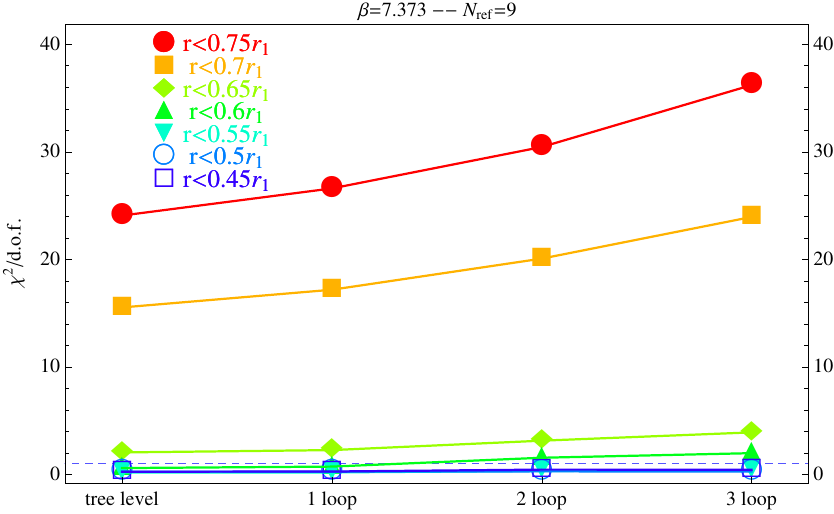}
\caption{Same as Fig.~\ref{fig:chisfits_beta7825} for $\beta=7.373$.}\label{fig:chisfits_beta7373}
\end{figure}

We perform the fits at tree-level, one-, two-, and three-loop accuracy. The reduced $\chi^2$ of the fits, for the $\beta=7.373, 7.596, 7.825$ sets, in each of our distance ranges are shown in Figs.~\ref{fig:chisfits_beta7825}-\ref{fig:chisfits_beta7373}.
From these figures we see that in the largest distance range we considered, i.e. $r<0.75r_1$, the reduced $\chi^2$s do not decrease when we increase the perturbative order. When we reduce the distance range, the $\chi^2$ curves move in the correct direction in order to satisfy the requirements in Sec.~\ref{subsec:procextr}. For $r<0.6r_1$ and shorter distance ranges, in all cases (with the only exception of the case $\beta=7.596$ with $N_{\rm ref}=9$), the $\chi^2$ curves are either almost flat (i.e. if they increase it is by one unit or less), or are at $\chi^2/d.o.f.\lesssim1$. Therefore, these distance ranges satisfy our conditions in Sec.~\ref{subsec:procextr} and are adequate for the $\als$ determination.

\begin{figure}
\centering
\includegraphics[width=8.6cm]{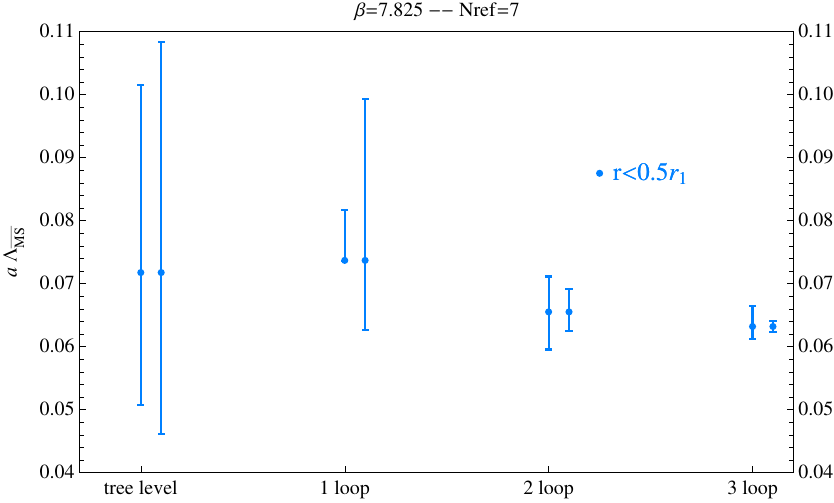}
\includegraphics[width=8.6cm]{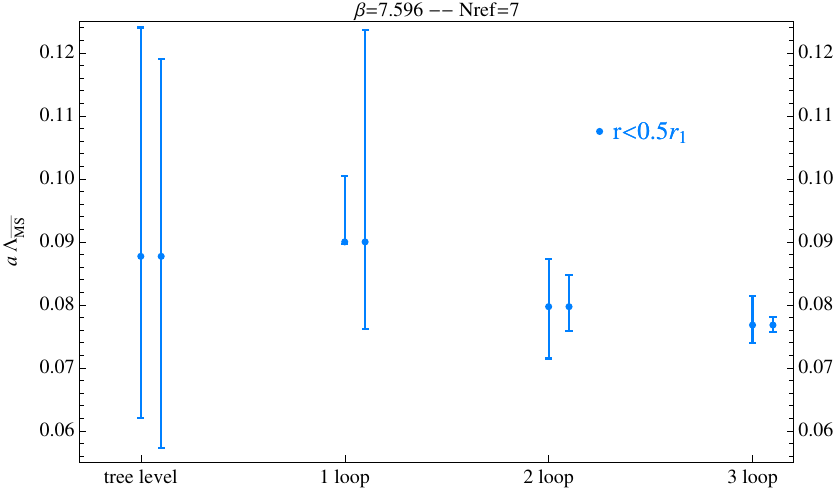}
\includegraphics[width=8.6cm]{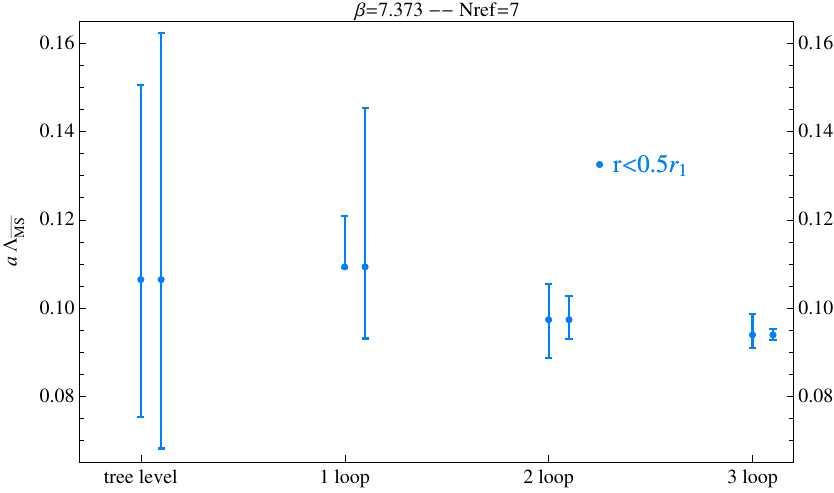}
\caption{Results for $a\Lambda_{\MS}$ from the fits to the lattice data at different orders of perturbative accuracy. For each order, the error bars of the point to the left correspond to the $\nu$ variation, while the ones of the point to the right correspond to the addition of a generic higher-order term, see Sec.~\ref{subsec:procextr}. The three panels correspond to the three different lattice spacings.}\label{fig:lams_rp5}
\end{figure}

For illustration, we show the results we obtain for $a\Lambda_{\MS}$ from the $r<0.5r_1$ distance range, at different orders of perturbative accuracy, in Fig.~\ref{fig:lams_rp5}. The figure is for $N_{\rm ref}=7$, but the other cases look quite similar. The error bars in the figure are obtained by repeating the fits, with the scale variation and addition of higher-order term, as specified in the previous section. For each perturbative order, the point to the left corresponds to the $\nu$ variation and the point to the right to the addition of a generic higher-order term. (We note that, for some numerical coincidence, the downward error bars from $\nu$ variation at one loop are very small). Both error bars reflect the perturbative uncertainty at a given order, and we take the largest of the two as our perturbative-uncertainty estimate. As we can clearly see in the figure, the central value at any given order is always nicely contained within the error bars of the previous order. This fact gives us confidence that the perturbative uncertainty is faithfully estimated. Note also the big error reduction in the determination of $a\Lambda_{\MS}$ that the knowledge of high-order terms brings about.

\begin{figure}
\centering
\includegraphics[width=8.6cm]{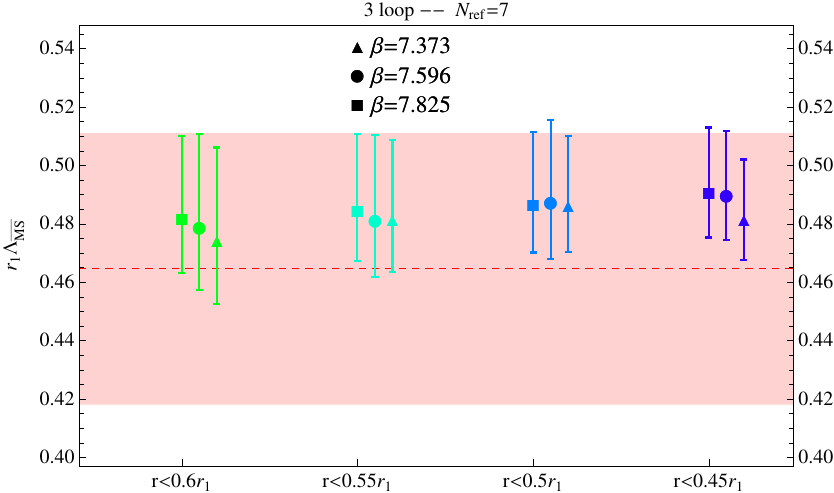}
\includegraphics[width=8.6cm]{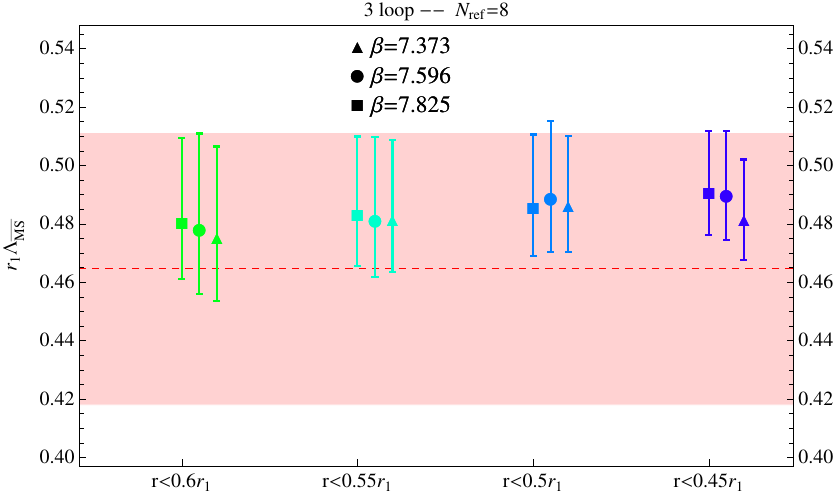}
\includegraphics[width=8.6cm]{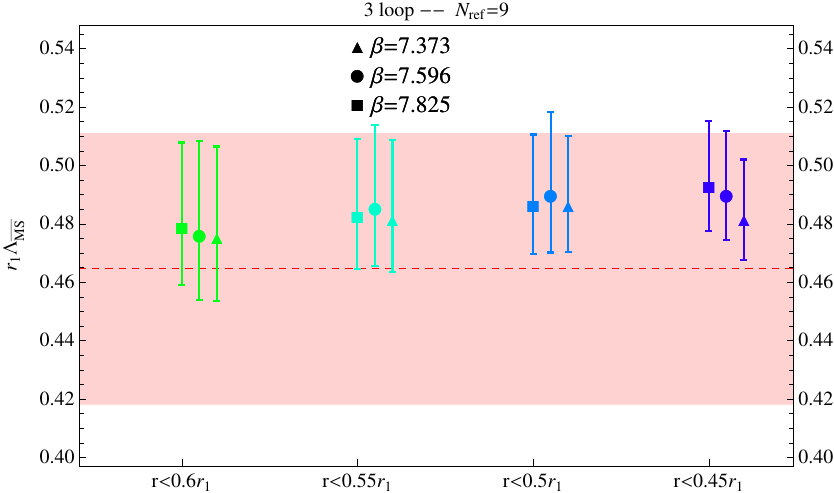}
\caption{Results for $r_1\Lambda_{\MS}$ at three-loop accuracy. The three panels correspond to different choices of $N_{\rm ref}$. For reference and comparison, the band shows our previous result in Ref.~\cite{Bazavov:2012ka}.}\label{fig:summlams}
\end{figure}

We show in Fig.~\ref{fig:summlams} the results we obtain for $r_1\Lambda_{\MS}$ for the $r<0.6r_1$ and shorter distance ranges, which are the ones that passed our criteria in Sec.~\ref{subsec:procextr}, at three-loop accuracy. We converted the results for $a\Lambda_{\MS}$ for each lattice spacing to $r_1\Lambda_{\MS}$ using the corresponding value of $r_1/a$ and its error, see Sec.~\ref{sec:lattice}; we added the perturbative error and the error in $r_1/a$ in quadrature in the error bars of Fig.~\ref{fig:summlams}. We note that the numbers obtained for the different distance ranges are perfectly compatible with each other. We will use the numbers for $r<0.5r_1$ in our final results, since the fits in this distance range always have reduced $\chi^2$s around 1. The effect of changing $N_{\rm ref}$ produces very small variations in the result, compare the three panels in Fig.~\ref{fig:summlams}. We use in any case the range spanned by our three choices of $N_{\rm ref}$, at each lattice spacing, to determine the corresponding result for $a\Lambda_{\MS}$. Then we can perform a weighted average of the results at the three different lattice spacings to obtain our final number for $r_1\Lambda_{\MS}$. Finally we convert it to physical units using the value of $r_1$ given in Sec.~\ref{sec:lattice}. All these numbers and our final result for $\als$ are collected in Sec.~\ref{subsec:asresult} below. In Fig.~\ref{fig:summlams} we also displayed, for comparison, our previous result in Ref.~\cite{Bazavov:2012ka} as the pink band. We can see that the present results are perfectly compatible with Ref.~\cite{Bazavov:2012ka}, and have a smaller error. 

\subsubsection{Statistical errors of the fit parameters}\label{subsubsec:staterr}
\begin{figure}
\centering
\includegraphics[width=8.6cm]{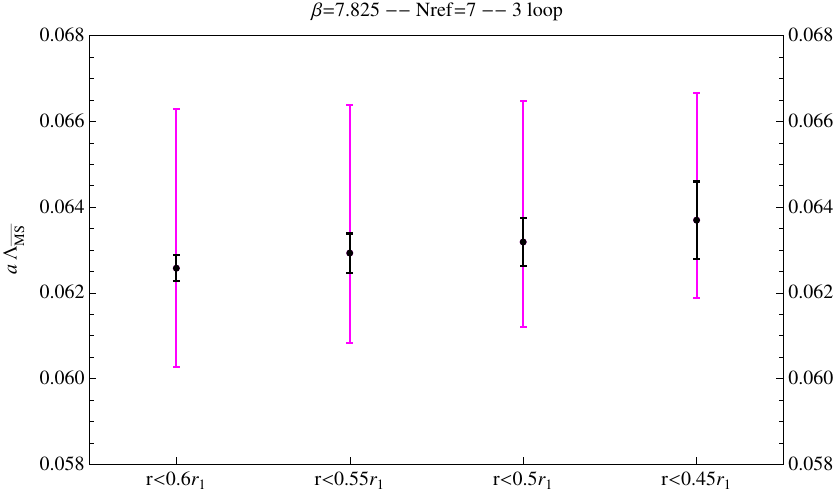}
\includegraphics[width=8.6cm]{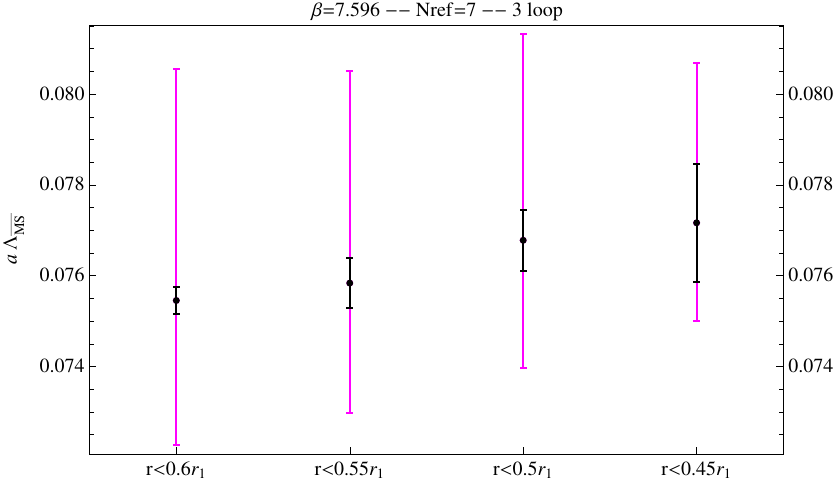}
\includegraphics[width=8.6cm]{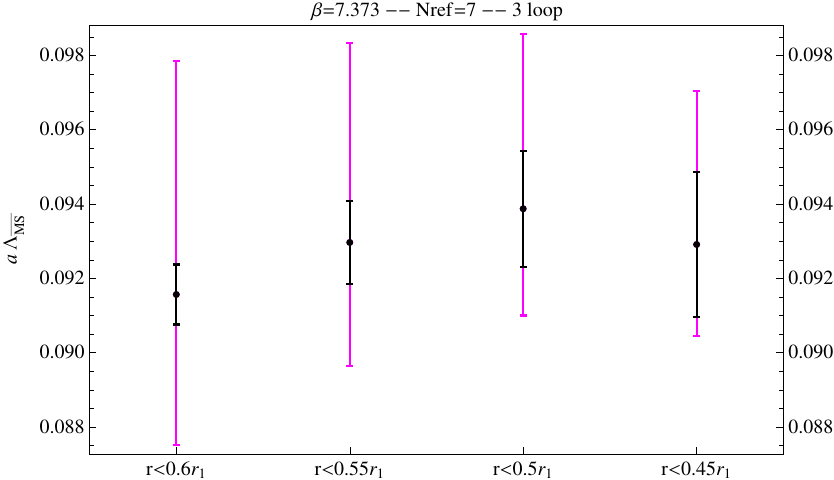}
\caption{Values for $a\Lambda_{\MS}$ with statistical (black -darker- error bars) and perturbative (magenta -lighter- error bars) errors for the different distance ranges. The three panels correspond to the three different lattice spacings.}\label{fig:staterr}
\end{figure}

In addition to the perturbative errors due to unknown higher-order terms (and the error due to the uncertainty in the values of $r_1/a$) that we showed above, there is also an uncertainty in the values of $a\Lambda_{\MS}$ due to the fit parameter errors. We estimate these errors, which we call statistical, by taking the values of $\Lambda_{\MS}$ at one $\chi^2$ unit above the minimum. Figure~\ref{fig:staterr} shows these statistical errors for the different distance ranges we consider (black -darker- error bars), together with the perturbative errors (magenta -lighter- error bars) for reference. The figure is for $N_{\rm ref}=7$, but the other cases look quite similar. We can see that the statistical errors are always smaller than the perturbative ones. They are not completely negligible, though, as was the case in our previous analysis in Ref.~\cite{Bazavov:2012ka}. 
We include these statistical errors in our final tables and results in Sec.~\ref{subsec:asresult}. We also note that the statistical errors for the $r<0.5r_1$ ranges are comparable in size with the errors due to the uncertainty in $r_1/a$. The final tables in Sec.~\ref{subsec:asresult} detail each one of the errors individually, and one can easily compare their relative sizes.

\subsubsection{Ultrasoft resummation}\label{subsubsec:usresan}
As mentioned in Sec.~\ref{subsec:resus}, the ultrasoft, i.e. $\ln\als(1/r)$, terms have been resummed at sub-leading accuracy. The numerical importance of these terms, and the necessity or not to include their resummation, depends on the distance $r$ we are studying. It is clear that their size grows when we go to shorter distances, since $\als(1/r)\to0$. In this section we want to study their numerical size in the range of distances where we compare with lattice. Note that, since in principle we do not know if it is necessary or not to include ultrasoft resummation (and at which order), it is more suitable to perform the study in the previous section (testing if the lattice data has reached or not the perturbative regime) before including the ultrasoft resummed expressions, as we have presented. 

\begin{figure}
\centering
\includegraphics[width=8.6cm]{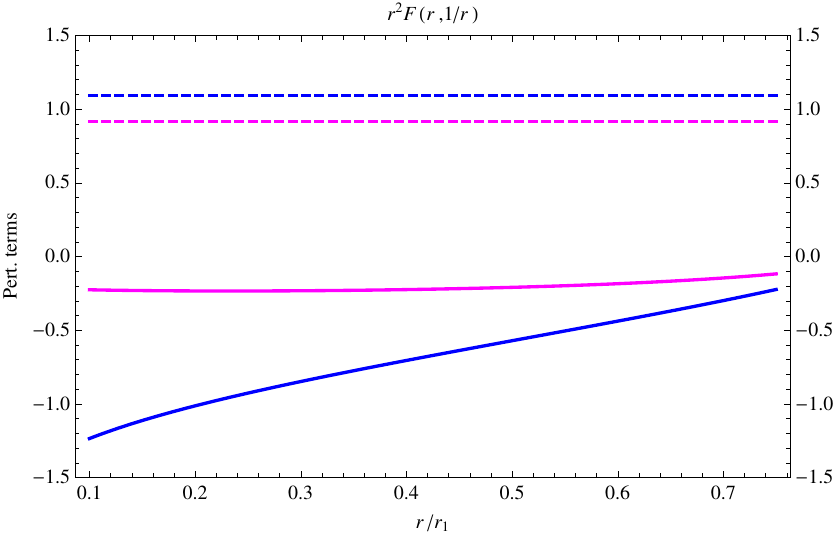}
\caption{Comparison of different perturbative terms for the force. The solid blue (darker) line corresponds to the ultrasoft logarithm at three loops, $(C_F/(4\pi)^3)a_3^L\ln(C_A\als/2)$, the dashed blue (darker) line is the rest of the three-loop term, i.e. $(C_F/(4\pi)^3)(\tilde{a}_3-6\tilde{a}_2\beta_0-4\tilde{a}_1\beta_1-2\beta_2)$. The dashed magenta (lighter) line is the two-loop term, $(C_F/(4\pi)^2)(\tilde{a}_2-4\tilde{a}_1\beta_0-2\beta_1)$, and the solid magenta (lighter) line corresponds to $(C_F\als/(4\pi)^3)a_3^L\ln(C_A\als/2)$. $r_1\Lambda_{\MS}=0.48$ is used for the solid curves (the dashed ones do not depend on $\als$).}\label{fig:checkus}
\end{figure}

To visualize the importance of the ultrasoft terms, we plot in Fig.~\ref{fig:checkus} the $a_3^L\ln(C_A\als/2)$ term (solid blue -darker- line), together with the rest of the three loop term for the force, i.e. $\tilde{a}_3-6\tilde{a}_2\beta_0-4\tilde{a}_1\beta_1-2\beta_2$ (dashed blue -darker- line), see Eq.~(\ref{eq:pertF}). From the figure we see that the ultrasoft log is much smaller than the non-logarithmic term for $r\sim0.75r_1$, but it grows quite rapidly as $r$ decreases; it becomes comparable to it for $r\sim0.2r_1$, and it is larger than the non-logarithmic part for the shortest distances where we have lattice data. We also notice that the ultrasoft logarithm and the non-logarithmic part have opposite signs, and largely compensate each other for distances around $0.2r_1$. Therefore, although the size of the ultrasoft logarithm at three loops is never much larger than the non-logarithmic part, it is certainly suitable to include leading-ultrasoft resummation in our analyses. Consequently, results including leading-ultrasoft resummation are also collected in Sec.~\ref{subsec:asresult} below. One would not expect, though, a big difference between the results at three loop with and without ultrasoft resummation.

In principle one could also argue that the leading ultrasoft resummation should be included together with the two-loop term, which would correspond to counting $\ln\als\sim1/\als$. Figure~\ref{fig:checkus} also shows the two-loop term $\tilde{a}_2-4\tilde{a}_1\beta_0-2\beta_1$ (dashed magenta -lighter- line), and compares it with $(\als/4\pi)a_3^L\ln(C_A\als/2)$ (solid magenta -lighter- line). We can see that the solid magenta line is never comparable in size with the two loop term. Therefore, in our present comparison with lattice, it is better not to include the leading ultrasoft resummation with the two-loop term. Instead it is more consistent to include it with the three-loop term, as we did in the previous paragraph. One could also consider including the sub-leading ultrasoft resummation in the analysis, since one of the reasons that motivated its computation was that the $\als^5\ln\als$ terms, computed in Ref.~\cite{Brambilla:2006wp}, were found to be quite large. Nevertheless, given the behavior of the three-loop terms shown in Fig.~\ref{fig:checkus}, it can not be excluded that a similar compensation, in the range of distances we are studying in this paper, between the four-loop $\als^5\ln\als$ terms and the unknown $\tilde{a}_4$ coefficient takes place. This means that if we were to include sub-leading ultrasoft resummation in the analysis, we would have to allow for generous perturbative uncertainties for the four-loop terms, to reflect for the aforementioned possible cancellations, and therefore the extraction of $\als$ would not really benefit much.

We have, in any case, performed the analyses with leading-ultrasoft resummation included along with the two-loop term (N$^2$LL accuracy), with leading-ultrasoft resummation included along with the three-loop term, and with sub-leading-ultrasoft resummation included along with the three-loop term (N$^3$LL accuracy). These studies allow us to verify the expectations described in the previous paragraph. What we find is that: (i) the N$^2$LL fits tend to give similar but a bit larger $\chi^2$ values than the corresponding two-loop fits, (ii) the fits with leading-ultrasoft resummation at three loops tend to give $\chi^2$ values that are very close to the three-loop ones, and (iii) the N$^3$LL fits tend to give larger $\chi^2$ values than the other fits with three-loop accuracy. This behavior is consistent with the observations in the previous paragraphs about the size of the ultrasoft terms. For these reasons, and as already mentioned above, we only present the numbers with leading-ultrasoft resummation included along with the  three-loop terms in our final results in Sec.~\ref{subsec:asresult}. We use the value $\mu=1.26r_1^{-1}\sim0.8$~GeV, for the ultrasoft factorization scale, in all our analyses. We have checked that variations of $\mu$ only produce small effects on the results, as could be expected since the impact of the ultrasoft resummation itself was already small, as we just discussed.

\subsection{Additional cross checks}\label{subsec:crosschecks}
Before presenting and collecting our final numbers in Sec.~\ref{subsec:asresult} below, we describe in this section several cross checks that we have performed. They serve to further verify the reliability of our procedure and results.

A first natural question is what would have happened if we had used our previous procedure in Ref.~\cite{Bazavov:2012ka} with the current enlarged lattice data set. We have repeated that analysis, and obtained $r_1\Lambda_{\MS}=0.48\pm0.05$ (at three-loop plus leading-ultrasoft-resummation accuracy). This number is perfectly compatible with the result quoted in Ref.~\cite{Bazavov:2012ka}, and has an error of similar size. The fact that the error did not really decrease, with respect to Ref.~\cite{Bazavov:2012ka}, may be seen as an indication that, to really benefit from the extended data set and the shorter distances reached, one should really employ expressions that completely avoid all $\ln(r\nu)$ terms, as we have done now. The previous figure is also compatible with the results obtained in the present paper with our modified analysis. 

\subsubsection{Comparison with lattice data for the force}\label{subsubsec:compforce}
One can also obtain lattice data for the force, and directly compare the perturbative expressions for the force with it. The problem in obtaining the lattice data for the force is that one needs to perform a numerical derivative of the lattice result for the energy, which is a non-trivial task. There are several uncertainties related to this numerical derivation, see Sec.~\ref{sec:lattice} for details. Since we need to introduce smoothing splines, and use synthetic data, to compute the derivative and obtain the force and its error, the resulting data we have for the force does not fluctuate independently. For this reason the $\chi^2$ analyses that we perform in our study may be seen as not completely adequate, or a bit unreliable. This fact led us to use the comparison with the energy, which does not suffer from this shortcoming, in our main analyses. It must nevertheless be checked what is the outcome if one compares with the force directly, which is what we do here.

\begin{figure}
\centering
\includegraphics[width=8.6cm]{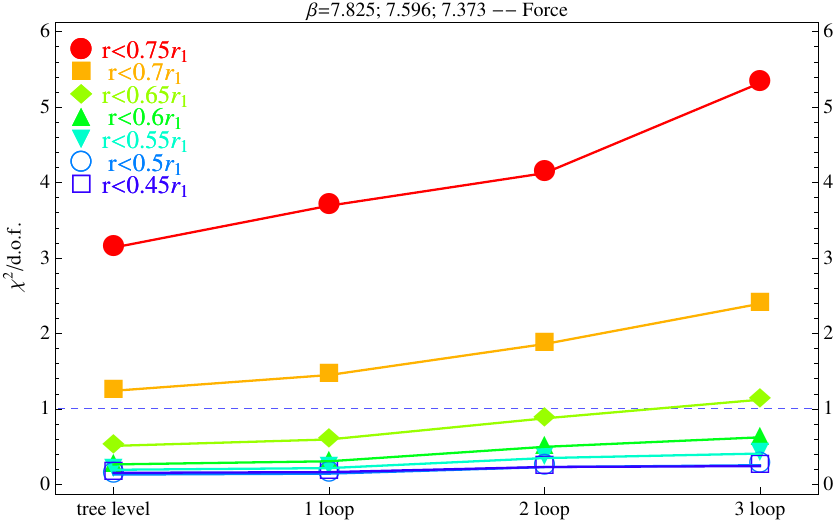}
\caption{Reduced $\chi^2$ of the fits at different orders of perturbative accuracy, and for several distance ranges, when comparing directly with the lattice data for the force. The dashed blue line marks $\chi^2/{\rm d.o.f.}=1$ for reference.}\label{fig:chis_forcelatt}
\end{figure}

\begin{figure}
\centering
\includegraphics[width=8.6cm]{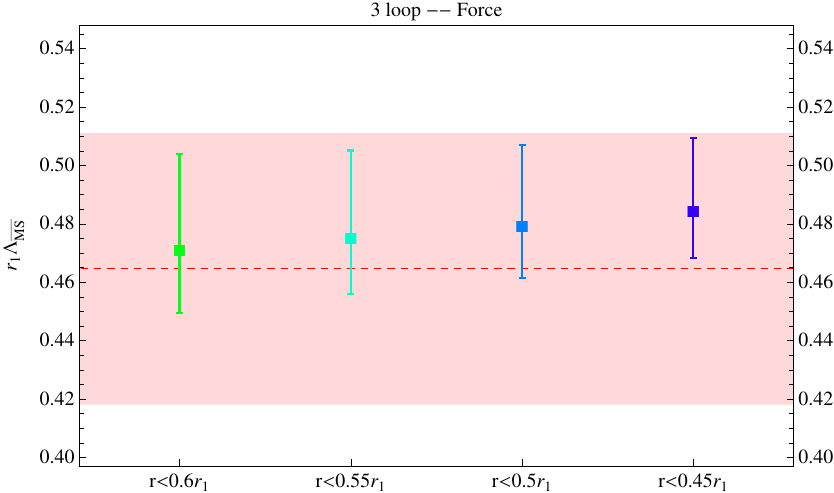}
\caption{Results for $r_1\Lambda_{\MS}$ at three-loop accuracy, when comparing directly with the lattice data for the force. For reference and comparison, the band shows our previous result in Ref.~\cite{Bazavov:2012ka}.}\label{fig:lams_forcelatt}
\end{figure}

To compare with the lattice data for the force, we can put the three data sets that we are using together, since the issue about the normalization, that affected the energy, is absent when we take the derivative. Apart from that, we follow the same procedure described in the previous sections, except that now we fit the perturbative expressions for $r^2F(r,1/r)$ to the lattice data for the force. In Fig.~\ref{fig:chis_forcelatt} we show the values of the reduced $\chi^2$s for the different distance ranges. As we can see from the figure, for distances $r<0.6r_1$, which are the ones that passed our criteria when doing the analyses with the energy in the previous sections, the $\chi^2$s are below one, confirming therefore that we can perform the extraction of $\als$ in these distance ranges. Actually, from Fig.~\ref{fig:chis_forcelatt}, it would seem that an even larger distance range, something like $r<0.7r_1$, could be used.
The results we obtain for $r_1\Lambda_{\MS}$ at three-loop accuracy, for $r<0.6r_1$ and shorter distance ranges, are shown in Fig.~\ref{fig:lams_forcelatt}. As we can see, the results are in very good agreement with what we obtained in the previous sections, compare with Fig.~\ref{fig:summlams}.

This agreement is an important cross check of the result, since, we recall again that, it is only the slope of $E_0(r)$, which is what the force encodes, that matters for the $\als$ extraction. We could have used this force data for our main analysis, but we preferred to leave it as a cross check because of the issues discussed above. 

\subsubsection{Analyses without lattice points with larger systematic uncertainties}\label{subsubsec:wosdp}
\begin{figure}
\centering
\includegraphics[width=8.6cm]{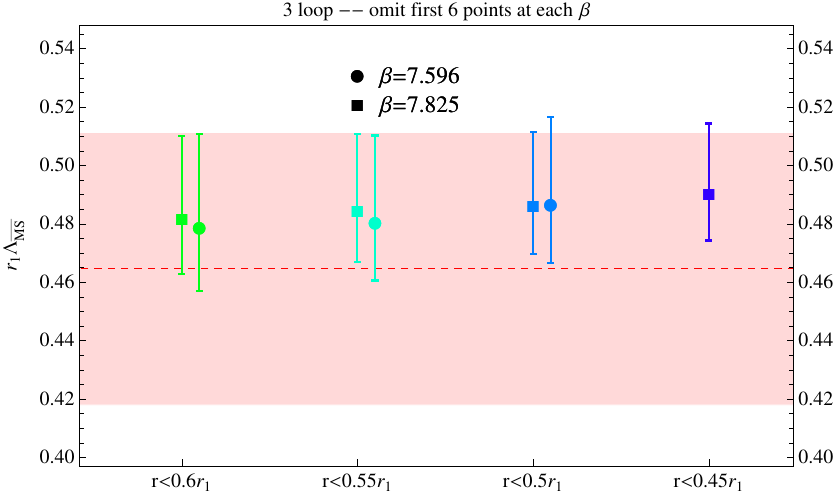}
\caption{Results for $r_1\Lambda_{\MS}$ at three-loop accuracy, when omitting the first six points for each lattice spacing. For reference and comparison, the band shows our previous result in Ref.~\cite{Bazavov:2012ka}.}\label{fig:lams_omit6}
\end{figure}

The lattice points for $E_0(r)$ at the shortest distances have larger discretization errors, which is part of the reason why we could not obtain reliable lattice data for the force at these distances. The way we estimate these errors was described in detail in Sec.~\ref{sec:lattice}. As explained there, our error estimates for these points should correctly reflect the size of the residual discretization errors in the data. Nevertheless, the points at larger distances, where the discretization errors are negligible, could be seen as more reliable than the ones at the shortest distances. With that in mind, we have repeated our analyses but dropping the first six points for each lattice spacing, which are the ones with larger systematic errors. We use $N_{\rm ref}=7$ for this analysis. We find that the $\chi^2$ curves present a behavior similar to that in Sec.~\ref{subsec:outana}, in particular the region $r<0.6r_1$ is suitable for the $\als$ extraction. We show the results for $r_1\Lambda_{\MS}$ at three-loop accuracy obtained with this reduced data set in Fig.~\ref{fig:lams_omit6}.
Note that we can only perform the fits when there are at least three data points. In particular no fits with the $\beta=7.373$ data set are possible in the $r<0.6r_1$ region, see Fig.~\ref{fig:dataranges}. The results in Fig.~\ref{fig:lams_omit6} are compatible and in good agreement with those of the previous sections. This shows that the inclusion of the points with larger systematic errors does not distort the result for $\als$, and can be seen as an indication that our estimation of the discretization errors is reliable.

\subsubsection{Analyses with only points at shorter distances}\label{subsubsec:onlysdp}
\begin{figure}
\centering
\includegraphics[width=8.6cm]{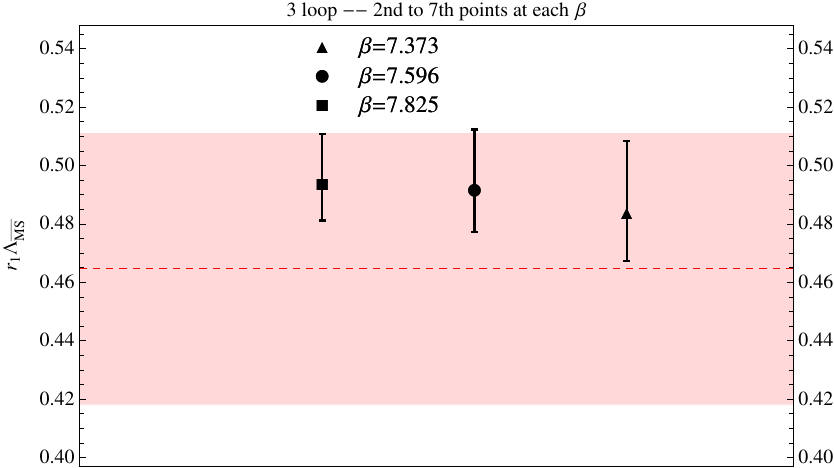}
\caption{Results for $r_1\Lambda_{\MS}$ at three-loop accuracy, using only the second to seventh point for each lattice spacing. For reference and comparison, the band shows our previous result in Ref.~\cite{Bazavov:2012ka}.}\label{fig:lams_2to7}
\end{figure}

One can also take the opposite point of view, with respect to the preceding section, and take for granted that our estimation of the systematic errors correctly reflects the residual discretization uncertainty. Then, the points at shortest distances should be the best ones for the $\als$ extraction, since this is the most perturbative region. With this view in mind, we repeated the analyses but now including only the second to seventh point at each lattice spacing; note that we chose to still omit the first point at each lattice spacing, which corresponds to a distance $r=a$, since for this point the reliability of the error estimate may be more questionable. We use $N_{\rm ref}=2$ for this analysis. There is only one distance range for each data set to perform fits in this case, i.e. that spanned by the second to seventh points. We find that the $\chi^2$s in this case are always below 1, which is perhaps not too surprising, since the errors for these points are larger than for the rest. We show in Fig.~\ref{fig:lams_2to7} the values of $r_1\Lambda_{\MS}$ obtained at three-loop accuracy in this case.
As one can see from the figure, the results are compatible with those in the previous sections. Note that, in particular, they are compatible with the results in Sec.~\ref{subsubsec:wosdp} above, where we did the opposite than here, and omitted the first six points at each lattice spacing.

\subsubsection{Analyses with extended distance ranges}

\begin{figure}
\centering
\includegraphics[width=8.6cm]{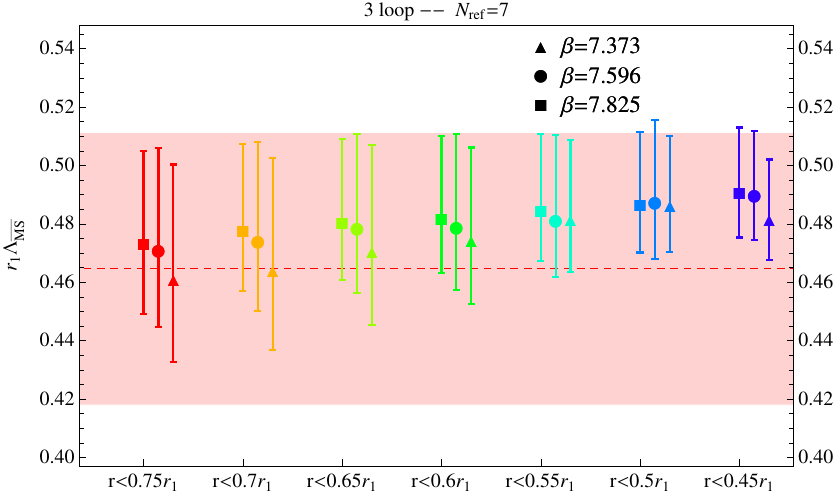}
\caption{Results for $r_1\Lambda_{\MS}$ at three-loop accuracy, also showing the outcome of analyses with extended distance ranges. For reference and comparison, the band shows our previous result in Ref.~\cite{Bazavov:2012ka}.}\label{fig:lams_ext_range}
\end{figure}

For completeness, in this section we present the results one obtains when using larger distance ranges in the fits, up to $r<0.75r_1$. Recall that distances $r<0.6r_1$ are the ones that passed our $\chi^2$ criteria, and were therefore deemed as suitable for the $\als$ extraction. The point of showing here the results from larger distance ranges is to illustrate that nothing dramatic happens beyond that point. Figure~\ref{fig:lams_ext_range} shows the results for $r_1\Lambda_{\MS}$ at three-loop accuracy, in all the distance ranges we have analyzed.
As one can see from the figure, the fits that use distances larger than $0.6r_1$ give results for $r_1\Lambda_{\MS}$ that are compatible with those used in our main analysis. The error bars, which, remember, come from unknown higher-perturbative orders, are larger in the extended distance ranges. This may be attributed to the fact that those fits involve lower-energy scales and therefore larger values of~$\als$. 

\subsubsection{Analyses of possible influence from non-perturbative-condensate terms}\label{subsubsec:condeff}
\begin{figure}
\centering
\includegraphics[width=8.6cm]{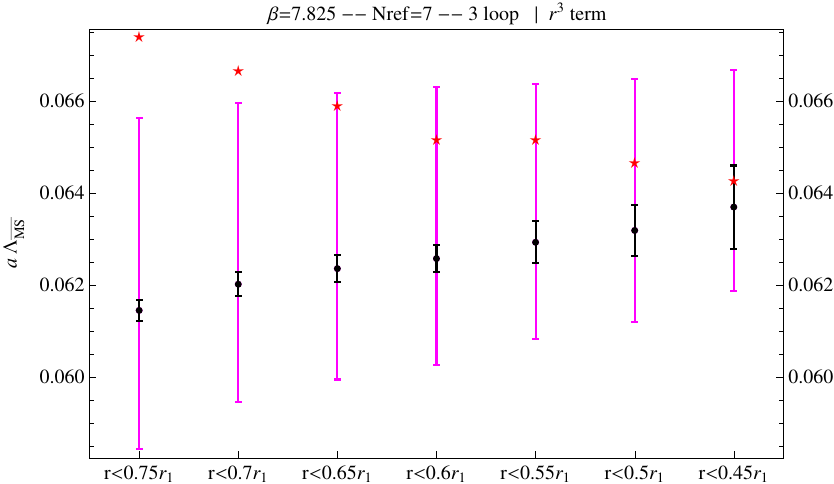}
\caption{Values for $a\Lambda_{\MS}$ from the fits including an $r^3$ monomial, red stars (see text), together with our default results with statistical (black -darker- error bars) and perturbative (magenta -lighter- error bars) errors.}\label{fig:fitr3}
\end{figure}

As discussed in previous sections, our $\chi^2$ criteria made manifest that we can safely use perturbation theory to describe the lattice data in the $r<0.6r_1$ distance ranges. Consequently, we used a purely perturbative expression in all our fits, and neglected any possible, parametrically suppressed, non-perturbative contributions to the static energy at these distances. Nevertheless, one might ask whether or not the presence of some non-perturbative term, which should necessarily be small, could distort the outcome of our fits in a significant way. To quantitatively address this question, we repeated the fits adding a monomial term, with a coefficient to be fitted, to our perturbative expression at three-loop accuracy. We considered $r^3$ and $r^2$ monomials, which could be associated with gluon and quark local condensates, and also an $r$ monomial. Our new fits contain therefore two parameters, which are $\Lambda_{\MS}$ and the coefficient of the monomial term. From the outcome of these fits, we do not find any evidence for a significant non-perturbative term in any of the cases. Furthermore, the values we obtain for $\Lambda_{\MS}$, at the $r<0.6r_1$ distance ranges, in these modified fits are perfectly consistent with the outcome of our previous fits. For illustration, we show the results for $\Lambda_{\MS}$ from the fits with an $r^3$ monomial, together with the outcome of our default fits, in Fig.~\ref{fig:fitr3}. The red stars in the figure are the results from the fit including the $r^3$ term, and the points with error bars are the results from our default fits.
We can clearly see from the figure that adding a monomial term, that could encode non-perturbative effects, does not distort our results for $\Lambda_{\MS}$. This fact provides further evidence that our extraction of $\als$ is robust.

It is worth stressing again that the goal of this section was to check that our perturbative analyses at short distances are stable and robust. A different question is how should one include non-perturbative effects to the perturbative expressions, in the form of local condensates, non-local condensates, etc., in order to extent their range of validity to larger distances. This is a very interesting question in itself, but it does not fall within the scope of the present paper, which is obtaining a precise determination of $\als$ from the short-distance region of the static energy. We leave these studies for future work.

\subsection{Final result for $\als$}\label{subsec:asresult}
In this section we collect the numbers we obtained from the analyses described in Secs.~\ref{subsec:procextr},~\ref{subsec:outana} and give our final result for $\als$. 

Table~\ref{tab:lam3l} summarizes the results we obtain at three-loop accuracy, for the $r<0.5r_1$ fit range. For each value of the lattice spacing, we present the result for $a\Lambda_{\MS}$, for each of the three points we use to define $N_{\rm ref}$, in the second to fourth columns in the table. As described in the previous sections, we take the range spanned by these three $N_{\rm ref}$ choices to get our result for $a\Lambda_{\MS}$ at each lattice spacing. This is shown in the fifth column of the table, the central value here is the average of the three previous columns. We then convert the result to $r_1$ units using the values of $r_1/a$ given in Sec.~\ref{sec:lattice}, and display it in the sixth column of the table. In all cases, the first error is the perturbative one, the second error is the one from statistical uncertainties in the fit, and, for the sixth column, the third error corresponds to the one coming from the value of $r_1/a$. We added the errors in quadrature on the right-hand side of the sixth column. Finally, we take a weighted average, with the inverse of the total error as the weight, of the results for the three different $\beta$ values to obtain our final number for $r_1\Lambda_{\MS}$. We added the errors linearly in this average, since the error at each $\beta$ value is dominated by the perturbative uncertainties, which are common to all data sets. This final number is shown on the bottom-right corner of the table.
\begin{table*}
\begin{tabular}{|c|c|c|c||c|c|}
\hline
& $a\Lambda_{\MS}$; $N_{\rm ref}=7$ &  $a\Lambda_{\MS}$; $N_{\rm ref}=8$ & $a\Lambda_{\MS}$; $N_{\rm ref}=9$ & $a\Lambda_{\MS}$; range spanned & $r_1\Lambda_{\MS}$; range spanned\\
\hline
$\beta=7.373$ & $0.0939^{+0.0047}_{-0.0029}$ & $0.0939^{+0.0047}_{-0.0029}$  & $0.0939^{+0.0047}_{-0.0029}$  &  $0.0939^{+0.0047}_{-0.0029}$ & $0.4855^{+0.0244}_{-0.0149}\pm0.0081\pm0.0024$\\
&$\pm0.0016$ &$\pm0.0016$ &$\pm0.0016$ & $\pm0.0016$ & $=0.4855^{+0.0258}_{-0.0171}$\\
\hline
$\beta=7.596$ & $0.0768^{+0.0046}_{-0.0028}$ & $0.0770^{+0.0043}_{-0.0027}$  & $0.0772^{+0.0046}_{-0.0029}$  & $0.0770^{+0.0048}_{-0.0030}$ & $0.4877^{+0.0302}_{-0.0191}\phantom{}^{+0.0059}_{-0.0058}\pm0.0043$\\
&$\pm0.0007$ &$\pm0.0009$ &$\pm0.0007$ & $\pm0.0009$ & $=0.4877^{+0.0311}_{-0.0204}$\\
\hline
$\beta=7.825$  & $0.0632^{+0.0033}_{-0.0020}$ & $0.0631^{+0.0033}_{-0.0020}$  & $0.0631^{+0.0033}_{-0.0020}$ & $0.0631^{+0.0034}_{-0.0021}$  & $0.4854^{+0.0258}_{-0.0160}\phantom{}^{+0.0061}_{-0.0062}\pm0.0037$\\
&$\pm0.0005$ &$\pm0.0007$ &$\pm0.0008$ & $\pm0.0008$ & $=0.4854^{+0.0268}_{-0.0175}$\\
\hline\hline
\textbf{Average}&\multicolumn{3}{c}{}&&\boldmath{$r_1\Lambda_{\MS}=0.486^{+0.028}_{-0.018}$}\\\hline
\end{tabular}
\caption{Results at three-loop accuracy for the $r<0.5r_1$ fit range. The second to fourth columns display the results for $a\Lambda_{\MS}$ ($a$ is the lattice spacing) for our three choices of $N_{\rm ref}$, and the fifth column is the range spanned by these values. This number is converted to units of $r_1$ in the sixth column. In all cases, the first error is the perturbative one, the second error is the statistical one, and, for the sixth column, the third error corresponds to the one coming from the value of $r_1/a$. The errors are added in quadrature on the right-hand side of the sixth column. The final average of the results for the three different $\beta$ values we use is given in the bottom-right corner of the table. See text for additional explanations.}\label{tab:lam3l}
\end{table*}

As explained in Sec.~\ref{subsubsec:usresan}, we also include here the results we obtain at three-loop plus leading-ultrasoft-resummation accuracy. These results are shown in Tab.~\ref{tab:lam3llusres}, with the same format as the previous table.
\begin{table*}
\begin{tabular}{|c|c|c|c||c|c|}
\hline
& $a\Lambda_{\MS}$; $N_{\rm ref}=7$ &  $a\Lambda_{\MS}$; $N_{\rm ref}=8$ & $a\Lambda_{\MS}$; $N_{\rm ref}=9$ & $a\Lambda_{\MS}$; range spanned & $r_1\Lambda_{\MS}$; range spanned\\
\hline
$\beta=7.373$ & $0.0957^{+0.0046}_{-0.0028}$ & $0.0957^{+0.0046}_{-0.0028}$  & $0.0957^{+0.0046}_{-0.0028}$  &  $0.0957^{+0.0046}_{-0.0028}$ & $0.4949^{+0.0240}_{-0.0144}\pm0.0086\pm0.0025$\\
&$\pm0.0017$ &$\pm0.0017$ &$\pm0.0017$ & $\pm0.0017$ & $=0.4949^{+0.0256}_{-0.0170}$\\
\hline
$\beta=7.596$ & $0.0781^{+0.0046}_{-0.0029}$ & $0.0784^{+0.0043}_{-0.0027}$  & $0.0785^{+0.0046}_{-0.0029}$  & $0.0783^{+0.0048}_{-0.0031}$ & $0.4961^{+0.0303}_{-0.0197}\phantom{}^{+0.0066}_{-0.0061}\pm0.0044$\\
&$\pm0.0007$ &$\pm0.0010$ &$\pm0.0007$ & $\pm0.0010$ & $=0.4961^{+0.0313}_{-0.0211}$\\
\hline
$\beta=7.825$  & $0.0644^{+0.0032}_{-0.0019}$ & $0.0642^{+0.0033}_{-0.0020}$  & $0.0643^{+0.0032}_{-0.0020}$ & $0.0643^{+0.0033}_{-0.0021}$  & $0.4944^{+0.0256}_{-0.0159}\pm0.0065\pm0.0037$\\
&$\pm0.0006$ &$\pm0.0008$ &$\pm0.0008$ & $\pm0.0008$ & $=0.4944^{+0.0267}_{-0.0175}$\\
\hline\hline
\textbf{Average}&\multicolumn{3}{c}{}&&\boldmath{$r_1\Lambda_{\MS}=0.495^{+0.028}_{-0.018}$}\\\hline
\end{tabular}
\caption{Same as Tab.~\ref{tab:lam3l} but for the results at three-loop plus leading-ultrasoft resummation accuracy. The average on the bottom-right corner constitutes the final result of this paper for $r_1\Lambda_{\MS}$.}\label{tab:lam3llusres}
\end{table*}
As we can see, the final results for $r_1\Lambda_{\MS}$ in Tabs.~\ref{tab:lam3l} and~\ref{tab:lam3llusres} are quite similar, as we expected from the discussion in Sec.~\ref{subsubsec:usresan}. We take the number at three-loop plus leading ultrasoft resummation accuracy as our best and final result. The final figure for $r_1\Lambda_{\MS}$, which we recall is for $n_f=3$ light-quark flavors and uses lattice data in the 1.3~GeV~-~5.1~GeV energy range, is therefore
\begin{equation}\label{eq:resr1Lambda}
r_1\Lambda_{\MS}=0.495^{+0.028}_{-0.018}.
\end{equation}
We now convert this result to physical units by using the value $r_1=0.3106\pm0.0008\pm0.0014\pm0.0004~{\rm fm}=0.3106\pm0.0017$~{\rm fm}, obtained in Ref.~\cite{Bazavov:2010hj} from the pion decay constant $f_{\pi}$ (where we added the errors in quadrature on the right-hand side). We obtain
\begin{equation}\label{eq:resLambda}
\Lambda_{\MS}=314.5^{+17.6}_{-11.7}\pm1.7~{\rm MeV}=315^{+18}_{-12}~{\rm MeV},
\end{equation}
where the first error corresponds to the one in Eq.~(\ref{eq:resr1Lambda}), and the second to the value of $r_1$, we added the two errors in quadrature on the right-hand side. This value of $\Lambda_{\MS}$ gives
\begin{equation}\label{eq:resas1p5}
\als(1.5~{\rm GeV},n_f=3)=0.336^{+0.012}_{-0.008},
\end{equation}
which corresponds to
\begin{equation}\label{eq:resasMZ}
\als(M_Z,n_f=5)=0.1166^{+0.0012}_{-0.0008},
\end{equation}
where we used the \texttt{Mathematica} package \texttt{RunDec}~\cite{Chetyrkin:2000yt} to obtain the numbers in Eqs.~(\ref{eq:resas1p5})-(\ref{eq:resasMZ})\footnote{Four-loop running, with the charm quark mass equal to 1.6~GeV and the bottom quark mass equal to 4.7~GeV. More in detail, we first convert $\Lambda_{\MS}$ to $\als$ at the highest scale we used in the extraction, i.e. 5.1~GeV, then evolve this value of $\als$ down to 1.5~GeV with a fixed $n_f=3$ number of flavors, and then evolve up to the scale $M_Z$ including the decoupling relations at the quark thresholds. We would like to emphasize that the actual outcome of our analyses is Eq.~(\ref{eq:resLambda}). We quote $\als(M_Z)$ because it has become
customary for the sake of comparison with other approaches. We do not
include errors associated to the truncation of the beta function to
four loops in the current version of \texttt{RunDec}. The effects of higher-order terms in the running have been estimated, and are negligible with the current accuracies, but may become relevant if the precision of $\alpha_s(M_Z)$ can be eventually reliably reduced to the per mil level~\cite{Moch:2014tta}.}. Equation~(\ref{eq:resasMZ}) constitutes the main result of the paper. This number is perfectly compatible with our previous result in Ref.~\cite{Bazavov:2012ka}, which was $\als(M_Z)=0.1156^{+0.0021}_{-0.0022}$, and supersedes it.

\section{Discussion and comparison with other works}\label{sec:disc}
\begin{figure}
\centering
\includegraphics[width=8.6cm]{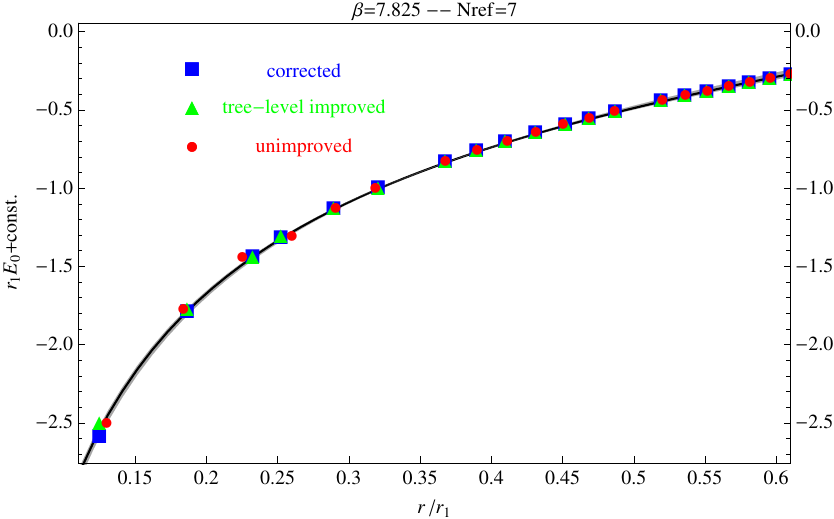}
\includegraphics[width=8.6cm]{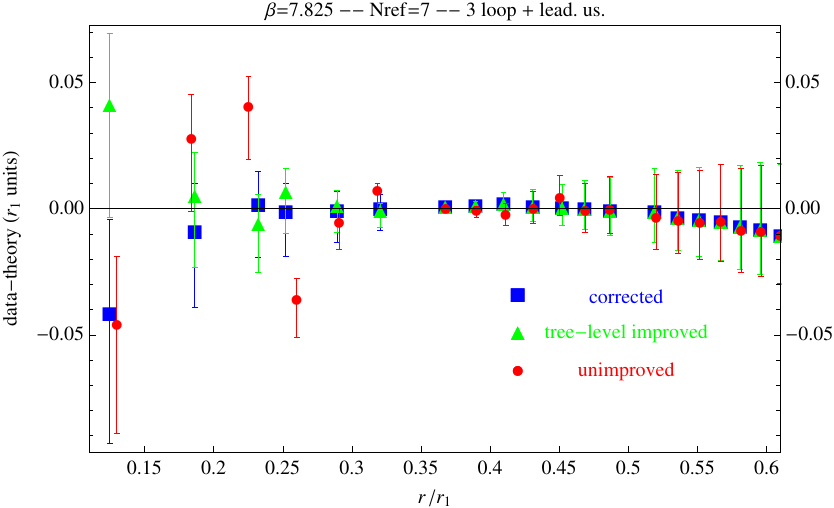}
\caption{Upper panel: Comparison of the lattice data for $\beta=7.825$ (blue squares) with the perturbative expression at three-loop plus leading ultrasoft resummation accuracy (black line). Our result in Eq.~(\ref{eq:resr1Lambda}), i.e. $r_1\Lambda_{\MS}=0.495^{+0.028}_{-0.018}$, is used for the perturbative expression, and the grey band corresponds to the variation of $r_1\Lambda_{\MS}$ within that interval. For illustration we also show the data points before correcting with the factors in Tab.~\ref{tab:corr} (green triangles), and before tree-level improvement (red circles), see text for additional explanations. Lower panel: Result of subtracting the perturbative expression from the lattice data. The error bars are obtained by adding, in quadrature, the errors of the lattice data and the uncertainty of the perturbative expression due to the variation of $r_1\Lambda_{\MS}$.}\label{fig:compbeta7825}
\end{figure}

\begin{figure}
\centering
\includegraphics[width=8.6cm]{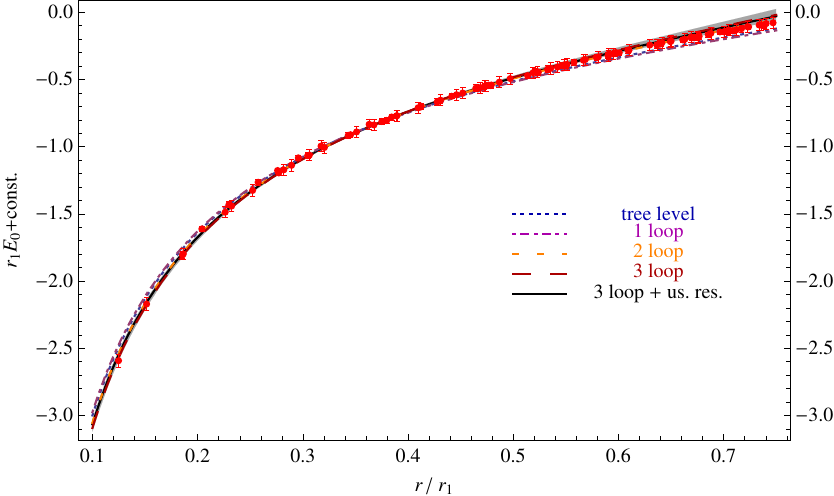}
\caption{Comparison of the lattice data for the static energy with perturbative expressions at different orders of accuracy. $r_1\Lambda_{\MS}=0.495$ is used for all the curves. The grey band corresponds to the variation $r_1\Lambda_{\MS}=0.495^{+0.028}_{-0.018}$ for the three-loop plus leading-ultrasoft-resummation accuracy curve. See text for additional explanations.}\label{fig:compallbeta}
\end{figure}

\begin{figure}
\centering
\includegraphics[width=8.6cm]{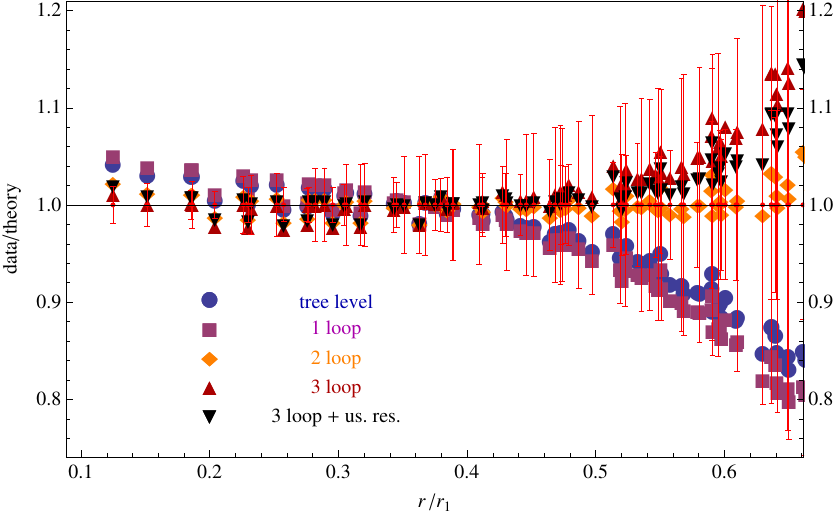}
\caption{Ratio of the lattice data over the theoretical prediction, at different orders of perturbative accuracy. $r_1\Lambda_{\MS}=0.495$ is used in all cases. The red error bars correspond to the errors of the lattice data. See text for additional explanations.}\label{fig:compdataratio}
\end{figure}

\begin{figure}
\centering
\includegraphics[width=8.6cm]{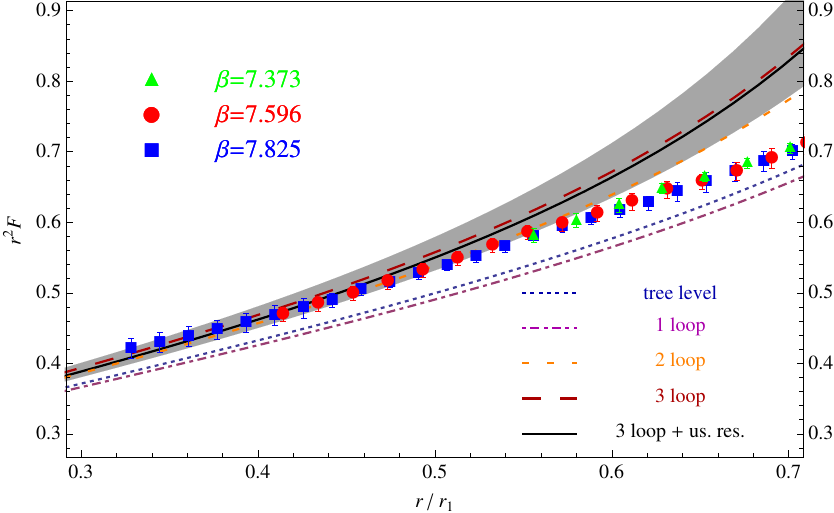}
\caption{Comparison of the lattice data for the force with perturbative expressions at different orders of accuracy. $r_1\Lambda_{\MS}=0.495$ is used for all the curves. The grey band corresponds to the variation $r_1\Lambda_{\MS}=0.495^{+0.028}_{-0.018}$ for the three-loop plus leading-ultrasoft-resummation accuracy curve. See text for additional explanations.}\label{fig:compforce}
\end{figure}

Having obtained our result for $\alpha_s$, we can now show how well the perturbative expressions describe the lattice data, when this number is used. To illustrate this, we show, in the upper panel of Fig.~\ref{fig:compbeta7825}, the lattice data for the smaller lattice spacing we used, i.e. $\beta=7.825$, (blue points) together with the perturbative prediction at three loop plus leading-ultrasoft resummation accuracy. The black curve corresponds to our central value for $\Lambda_{\MS}$, and the (barely visible in that scale) grey band is obtained by varying $\Lambda_{\MS}$ within our obtained range. Note that the plot is in units of $r_1$, but no error associated to normalization of the lattice data is shown, since that error is irrelevant when only one lattice set is used and the perturbative expression is normalized to it, as we do in this plot. For illustration, the figure also shows the lattice data before correcting it with the factors in Tab.~\ref{tab:corr} (green triangles), and, in addition to that, without using tree-level improvement (red circles). The error bars in these two last sets of data are statistical only, since the systematic errors due to discretization are estimated as part of the correction procedure. To see more clearly how well the theoretical expression describes the lattice data, we also show, in the lower panel of the figure, the result of subtracting the perturbative prediction from the lattice data. The error bars in that plot are obtained by adding in quadrature the errors of the lattice data and the band in the theoretical expression (which was due to the variation of $\Lambda_{\MS}$). Fig.~\ref{fig:compbeta7825} clearly shows that the perturbative expression can perfectly describe the corrected lattice data in that distance range.~(We recall that it is more difficult to reliably assess discretization errors for the point at shortest distance, see Sec.~\ref{sec:lattice}.)~In Fig.~\ref{fig:compallbeta} we put together the data for all the lattice spacings we have, including those used in Ref.~\cite{Bazavov:2012ka}, i.e. from $\beta=6.664$ to $\beta=7.825$, and compare them with the perturbative expressions at different orders of accuracy. The uncertainties due to the normalization of the lattice data to a common scale are now included in the error bars, as it is appropriate when putting together data from different lattice spacings. To visualize more easily the differences among the different perturbative orders, we also show in Fig.~\ref{fig:compdataratio} the ratio of the lattice data over the theoretical prediction. Finally, Fig.~\ref{fig:compforce} shows, for completeness, the direct comparison of the perturbative expressions with the lattice data for the force presented in Sec.~\ref{sec:lattice}.

It is worth stressing at this point that the knowledge of the static energy at three-loop accuracy was crucial in order to obtain our precise extraction of $\als$ in Eq.~(\ref{eq:resasMZ}). Although the fits at lower orders of perturbative accuracy can already give low $\chi^2$ values, see Figs.~\ref{fig:chisfits_beta7825}-\ref{fig:chisfits_beta7373}, the uncertainty that one should associate to a value of $\als$ extracted at lower orders is much larger. This was clearly illustrated in Fig.~\ref{fig:lams_rp5}, where the dramatic reduction of the error bars when increasing the perturbative accuracy of the prediction is manifest. We also recall that the different perturbative orders always refers to the corrections to the potential, i.e. the $\tilde{a}_i$ coefficients in Eq.~(\ref{eq:pertF}), but we always use four-loop accuracy for the running of $\als$. If one does not do that, and uses the running of $\als$ at lower accuracies, the $\chi^2$ values resulting from lower orders in perturbation theory are much higher.

\begin{figure}
\centering
\includegraphics[width=8.6cm]{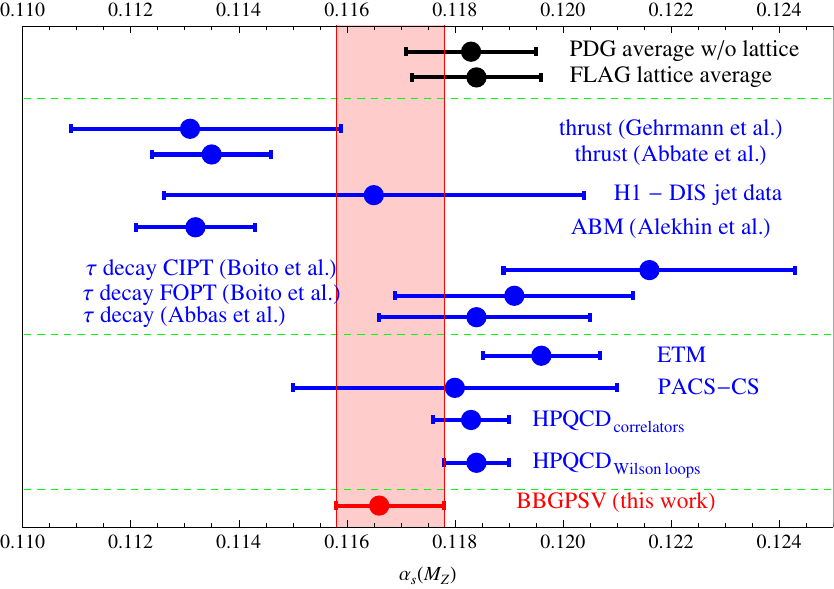}
\caption{Comparison of our result for $\als(M_Z)$ in Eq.~(\ref{eq:resasMZ}) -red point- with a few other recent determinations -blue points-. We include: The lattice determinations of the HPQCD~\cite{McNeile:2010ji}, PACS-CS~\cite{Aoki:2009tf} (we take the number quoted in Ref.~\cite{Aoki:2013ldr}), and ETM~\cite{Blossier:2013ioa} collaborations (the latter paper uses $n_f=2+1+1$ simulations while the other two use $n_f=2+1$ simulations). Non-lattice determinations from $\tau$ decays in Refs.~\cite{Abbas:2012fi,Boito:2012cr}; from thrust in $e^+e^-$ collisions in Refs.~\cite{Abbate:2010xh,Gehrmann:2012sc}; the recent H1 re-analysis of Ref.~\cite{Andreev:2014wwa}; and the PDF-fit ABM13 result of Ref.~\cite{Alekhin:2013nda} (note that the error bars in this case do not include effects from higher unknown perturbative orders). We also show the lattice average by the FLAG collaboration~\cite{Aoki:2013ldr} and the PDG average excluding lattice results~\cite{Beringer:1900zz} -black points-.}\label{fig:compalphas}
\end{figure}

Let us now compare our result for $\Lambda_{\MS}$ with other recent extractions of the strong coupling. The present analysis, together with our preceding paper~\cite{Bazavov:2012ka}, constitutes, at present, the only extraction of $\Lambda_{\MS}$ from the QCD static energy with at least three flavors; therefore, the only one that can be used to obtain $\als(M_Z)$. Other analyses aiming at extracting $\Lambda_{\MS}$ from the static energy with less than three flavors include Refs.~\cite{Leder:2011pz,Jansen:2011vv,Brambilla:2010pp,Bali:1992ru,Booth:1992bm,Sumino:2005cq,Karbstein:2014bsa}. In particular, let us mention that the $n_f=2$ analysis of Ref.~\cite{Leder:2011pz} concludes that smaller lattice spacings than those currently available to them would be needed to extract $\Lambda_{\MS}$. This reference uses Wilson fermions. The data for the force that they use contains only three points below $r=0.75r_1$, with none below $r=0.5r_1$. Our analyses show that we would not have been able to obtain $\Lambda_{\MS}$ with that amount of data. In this sense, we do agree with the findings of Ref.~\cite{Leder:2011pz}. Regarding lattice $\als$ extractions from other observables, the FLAG collaboration recently presented, in Ref.~\cite{Aoki:2013ldr}, a comprehensive and critical review of all the available $\als$ lattice determinations, and provided an average. We show this lattice average, together with our new result, in Fig.~\ref{fig:compalphas} (note that the FLAG average includes the result from Ref.~\cite{Bazavov:2012ka}). We also show in the figure a few other individual lattice determinations of $\als$, a few selected recent non-lattice determinations, and the PDG average excluding lattice results~\cite{Beringer:1900zz}. Further determinations of $\als$, as well as discussions about them, can be found, for instance, in the summary reports of recent dedicated workshops~\cite{Moch:2014tta,Bethke:2011tr}.

\section{Summary of results and conclusions}\label{sec:concl}
We have improved our previous extraction of $\alpha_s$, in Ref.~\cite{Bazavov:2012ka}, from the comparison of lattice data with perturbative expressions for the static energy of a heavy quark-antiquark pair. This has been possible because a considerable amount of new lattice data at shorter distances has become available~\cite{Bazavov:2014noa}, which has allowed us to carry out an extra number of cross checks and hence to considerably reduce the systematic errors. In particular we have been able to correct for cutoff effects in the shorter-distance points, to analyze the dependence on the fit range, and to carry out separate analyses for different lattice spacings. Thus we could, for instance, discard points which are not in the perturbative regime, points which suffer from large cutoff effects and, very importantly, avoid the lattice normalization errors that dominated our previous extraction. On the other hand, we have used improved perturbative expressions, in which not only the first renormalon is avoided, and the ultrasoft logarithms resummed, but also the soft logarithms are summed up. This appears to be necessary because the lattice data covers a relative large range of distance values now. Our final result reads
\begin{eqnarray}
r_1\Lambda_{\MS}=0.495^{+0.028}_{-0.018},
\end{eqnarray}
which corresponds to
\begin{eqnarray}
\als(M_Z,n_f=5) & = & 0.1166^{+0.0012}_{-0.0008}.
\end{eqnarray}
This updated result reduces the errors by roughly a factor of two with respect to our previous extraction. It displays a higher central value, which is, nevertheless, perfectly compatible with our previous result.

The errors of the $\alpha_s$ extraction presented here can in principle be reduced by just incorporating more lattice data at shorter distances, with no further modification of the perturbative expressions, which are already at the three-loop level. Notice that with the lattice data available at present, there is still little sensitivity to the ultrasoft resummation, and hence we do not expect much sensitivity to the yet unknown four-loop contribution. We have also checked that there is no sensitivity to other possible non-perturbative effects, like for instance those due to gluon or quark condensates.

We conclude that the method first outlined in Ref.~\cite{Brambilla:2010pp}, and further developed in the present paper, is not only able to produce competitive extractions of $\alpha_s$ when confronted with realistic lattice data~\cite{Bazavov:2012ka}, but also to consistently improve on the outcome upon the incorporation of new shorter-distance data. For the future, to further corroborate the result obtained here, it would be important to determine $\als$ from the static energy calculated using other lattice actions.

\begin{acknowledgments}
X.G.T. thanks Rainer Sommer for a useful conversation during a visit to DESY Zeuthen, which partially triggered the present updated analysis. This work was supported in part by U.S. Department of Energy under
Contract No. DE-AC02-98CH10886. The work of X.G.T. is supported by the Swiss National Science Foundation (SNF)
under the Sinergia grant number
CRSII2\underline{ }141847\underline{ }1. The work of N.B. and A.V. is supported in part by DFG and NSFC (CRC110) and they acknowledge
financial support from the DFG cluster of excellence ``Origin and
structure of the universe'' (www.universe-cluster.de). J.S. is supported by the CPAN CSD2007-00042 Consolider-Ingenio 2010 program (Spain), the 2009SGR502 CUR grant (Catalonia) and the FPA2010-16963 project (Spain). X.G.T. is grateful to the Mainz Institute for Theoretical Physics (MITP) for its hospitality and its partial support during the completion of this work.
\end{acknowledgments}

\end{document}